\documentclass{aa} 
\usepackage{txfonts}
\usepackage{graphicx}
\usepackage{subcaption}
\usepackage{natbib}
\usepackage{booktabs}
\usepackage{float}
\usepackage{color}
\usepackage{amsmath}
\usepackage{epstopdf}
\usepackage{textcomp}
\usepackage{gensymb}
\usepackage[colorlinks, citecolor={blue}]{hyperref}

\begin{document} 
	\title{Unveiling the environment and faint features of the isolated galaxy CIG\,96 with deep optical and HI observations}
   \titlerunning{Faint features of the isolated galaxy CIG\,96}

   \author{P.\,Ram\'irez-Moreta\inst{\ref{iaa}},
          L.\,Verdes-Montenegro\inst{\ref{iaa}},
          J.\,Blasco-Herrera\inst{\ref{iaa}},
          S.\,Leon\inst{\ref{eso}},
          A.\,Venhola\inst{\ref{kapteyn}, \ref{oulu}},
          M.\,Yun\inst{\ref{umass}},
          V.\,Peris\inst{\ref{val}},
	      R.\,Peletier\inst{\ref{kapteyn}},
	      G.\,Verdoes\,Kleijn\inst{\ref{kapteyn}},
	      E.\,Unda-Sanzana\inst{\ref{antof}},
	      D.\,Espada\inst{\ref{naoj1},\ref{naoj2}},
	      A.\,Bosma\inst{\ref{mars}},
		  E.\,Athanassoula\inst{\ref{mars}},
		  M.\,Argudo-Fern\'andez\inst{\ref{antof}},
		  J.\,Sabater\inst{\ref{edin}},
		  J.\,C.\,Muñoz-Mateos\inst{\ref{eso}},
		  M.\,G.\,Jones\inst{\ref{iaa}},
		  W.\,Huchtmeier\inst{\ref{mplanck}},
		  J.E.\,Ruiz\inst{\ref{iaa}},
          J.\,Iglesias-P\'{a}ramo\inst{\ref{iaa},\ref{almeria}},
          M.\,Fern\'andez-Lorenzo\inst{\ref{iaa}},
		  J.\,Beckman\inst{\ref{iac}},
          S.\,S\'anchez-Exp\'osito\inst{\ref{iaa}},
          J.\,Garrido\inst{\ref{iaa}}
	      }

   \institute{Instituto de Astrof\'isica de Andaluc\'ia (IAA-CSIC), Glorieta de la Astronom\'ia s/n, 18008, Granada, Spain\\\email{prm@iaa.es, pramirezmoreta@gmail.com}\label{iaa}
        \and Joint ALMA Observatory - ESO, Av. Alonso de Córdova, 3104 Santiago, Chile\ \label{eso}
		\and Kapteyn Instituut, Postbus 800, 9700 AV Groningen, The Netherlands\ \label{kapteyn}
		\and Astronomy Research Unit, University of Oulu, 90014, Oulu, Finland\ \label{oulu}
        \and Department of Astronomy, University of Massachusetts-Amherst (LGRT-B 522 710 North Pleasant Street, Amherst, MA, USA)\ \label{umass}
        \and Observatori Astron\`omic de la Universitat de Val\`encia, Catedr\'atico Jos\'e Beltr\'an, 2, 46980 Paterna, Spain\ \label{val}
        \and National Astronomical Observatory of Japan (NAOJ), 2-21-1 Osawa, Mitaka, 181-8588 Tokyo, Japan \ \label{naoj1}
        \and The Graduate University for Advanced Studies (SOKENDAI), 2-21-1 Osawa, Mitaka, 181-0015 Tokyo, Japan \ \label{naoj2}
        \and Aix$-$Marseille Universit\'e, CNRS, CNES, LAM, Marseille, France \ \label{mars}
        \and Institute for Astronomy, University of Edinburgh, Royal Observatory,
Blackford Hill, Edinburgh, EH9 3HJ, UK \ \label{edin}
		\and Universidad de Antofagasta, Unidad de Astronom\'ia, Facultad Cs. B\'asicas, Av. U. de Antofagasta 02800, Antofagasta, Chile\ \label{antof}
		\and Max-Planck-Institut für Radioastronomie, Auf dem Hügel 69, D-53121 Bonn, Germany \ \label{mplanck}
		\and Estaci\'{o}n Experimental de Zonas \'{A}ridas (CSIC), Ctra. de Sacramento s/n, La Ca\~{n}ada de San Urbano, 04120 Almer\'{\i}a, Spain \ \label{almeria}
		\and Instituto de Astrof\'isica de Canarias, c/V\'ia L\'actea, s/n, E-38205, La Laguna, Tenerife, Spain \ \label{iac}
			}
		     
   \authorrunning{Ram\'irez-Moreta et al.} 

   \date{Received Month Day, Year; accepted Month Day, Year}

  \abstract
   {Asymmetries in atomic hydrogen (HI) in galaxies are often caused by the interaction with close companions, making isolated galaxies an ideal framework to study secular evolution. The AMIGA project has demonstrated that isolated galaxies show the lowest level of asymmetry in their HI integrated profiles compared to even field galaxies, yet some present significant asymmetries. CIG\,96 (NGC 864) is a representative case reaching a 16\% level.}
   {Our aim is to investigate the HI asymmetries of the spiral galaxy CIG\,96 and what processes have triggered the star-forming regions observed in the XUV pseudo-ring.}
   {We performed deep optical observations at CAHA1.23m, CAHA2.2m and VST (OmegaCAM wide-field camera) telescopes. We reach surface brightness (SB) limits of $\mu_{CAHA2.2m}$\,=\,27.5\,mag\,arcsec$^{-2}$ (Cousins $R$) and $\mu_{VST}$\,=\,28.7\,mag\,arcsec$^{-2}$ (SDSS\,$r$) that show the XUV pseudo-ring of the galaxy in detail. Additionally, a wavelet filtering of the HI data cube from our deep observations with VLA/EVLA telescope allowed us to reach a column density of $N_{HI}$ = 8.9\,$\times$\,10$^{18}$\,cm$^{-2}$\,(5$\sigma$) (28$\arcsec$ $\times$ 28$\arcsec$ beam), lower than in any isolated galaxy.}
   {We confirm that the HI of CIG\,96 extends farther than 4$\times$\,r$_{25}$ in all directions. Furthermore, we detect for the first time two gaseous structures ($\sim$10$^{6}$\,M$_{\odot}$) in the outskirts. The SDSS $g-r$ colour index image from CAHA1.23m shows extremely blue colours in certain regions of the pseudo-ring where $N_{HI} >8.5\times10^{20}$\,cm$^{-2}$, whereas the rest show red colours. Galactic cirrus contaminate the field, setting an unavoidable detection limit at 28.5\,mag\,arcsec$^{-2}$ (SDSS\,$r$).}
   {At the current SB and $N_{HI}$ levels, we detect no stellar link within 1\degree$\times$1\degree or gaseous link within 40\arcmin$\times$40\arcmin\ between CIG\,96 and any companion. The isolation criteria rule out interactions with other similar-sized galaxies for at least $\sim$2.7\,Gyr. Using existing stellar evolution models, the age of the pseudo-ring is estimated at 1\,Gyr or older. Undetected previously accreted companions and cold gas accretion remain as the main hypothesis to explain the optical pseudo-ring and HI features of CIG\,96.}


   \maketitle


\section{Introduction}
\label{sec:intro}

Most galaxies in the nearby universe are either interacting with or gravitationally bound to nearby companions. Such events are directly responsible for a continuous change in their structural, dynamical and chemical properties \citep{toomre77}. A wide set of observed and, broadly, understood effects of such interactions (e.g. quenching or enhancement of the stellar formation, gaseous plumes and bridges, tidal streams, etc.) constitute some of the main drivers of the evolution of galaxies. 
Such interactions may prevail over the internal processes, hiding or even disrupting the key inner evolutionary mechanisms of each particular galaxy that, in the absence of large companions, might otherwise dominate its evolution. 
The bars present in nearly two$-$thirds of the spiral galaxies (e.g. \citealt{buta15}), whether initially from external or internal origin, are among these inner elements that can crucially affect the evolution of the galaxy from their bulges or pseudo-bulges out to the outer Lindblad resonance in their external regions \citep[e.g.][]{korken04, buta05, fernandezlorenzo14}.
Additionally, the results from other cosmologically motivated studies point out that the interaction of the galaxies with dark matter halos might result in perturbations of the disc (e.g. \citealt{kazantzidis08, kazantzidis09}).
Isolated galaxies, if selected with strict and robust criteria, constitute an ideal framework to study the secular evolution of galaxies since we can exclude the possibility of interactions with large companions.
The Analysis of the interstellar Medium of Isolated GAlaxies (AMIGA) project\footnote{\href{http://amiga.iaa.es/}{http://amiga.iaa.es}} \citep{verdes05} was designed to perform a multi-wavelength study of a large sample of galaxies selected with strict isolation criteria from the Catalog of Isolated Galaxies (CIG\footnote{This catalog is referred to as K73 in SIMBAD and KIG in NED databases.}, 1051 galaxies, \citealt{kar73}).

With respect to the isolation level, a plethora of references to different definitions and selection criteria may be found throughout the literature of the last 40 years, (e.g. all references in \citealt{verdes05} or \citealt{muldrew12} among others). As part of the AMIGA project, this work makes use of its isolation criteria and parameters (local number density $\eta_{k}$ and tidal force estimation $Q$) in the version by \cite{verley07b} revised later by \citealt[][]{argudo13, argudo14}. Both isolation parameters are defined in depth in the discussion of the environment (see Sect.\,\ref{sec:environment}).

The results of the project are that variables expected to be enhanced by interactions are lower in isolated galaxies than in any other sample (e.g. L$_{FIR}$, \citealt{lisenfeld07}, radio continuum emission, optical symmetry, \citealt{verdes10} and references therein, active galactic nucleus (AGN) rate, \citealt{sabater12}). Among these, one specific result is especially significant in the context of the present work: the asymmetry level of the atomic gas (HI) integrated profiles of the CIG galaxies is also lower than any other sample, including field galaxies (\citealt{espada11b}, see \citealt{jones18} for a full characterisation of the HI content of AMIGA sample). However, a number of galaxies show unusually high levels of asymmetry (up to 50\%), the causes of which remain unknown.


If asymmetries can only be generated by interactions, lopsidedness in an isolated galaxy such as CIG\,96 (NGC\,864) should not be observed. However, previous data from Green Bank as well as VLA observatories show a large HI envelope beyond 2$\times$ $r_{25}$ (q.v. Table\,\ref{table:cig96}) that has an asymmetry level of 16\% in its HI integrated profile \citep{espada05}.

\cite{espada11a} report on a partial XUV ring (hereafter the pseudo-ring, see Sect.\,\ref{sec:sblimit}) seen in near-UV (NUV) and far-UV (FUV) GALEX data, and located at 1.5$-$2$\times$ $r_{25}$. This pseudo-ring shows patchy regions with star formation (SF). It is not clear that such features can develop in galaxies free from interactions. In this paper we present additional data on this enigmatic object, in particular by obtaining further deep imaging at optical wavelengths.


\cite{erroz12} have studied the kinematics of the inner regions of CIG\,96 in H$\alpha$ but no previous study has provided convincing arguments that an external agent can explain both the HI and optical features of CIG\,96. As a consequence, this raises the question as to whether asymmetries might develop in galaxies free from interactions \citep{espada05, espada11a}, motivating the in-depth study of CIG\,96.

However, to support any internal agent as the main evolutionary process, it is necessary to first rule out any external influence. Neither tidal features nor gas$-$rich companions are found in HI maps even for the most asymmetric cases \citep[e.g.][]{espada05, espada11b, portas11, sengupta12} and current shallow optical images are surprisingly symmetric when dust patches are ignored. In the absence of interactions for the last $\sim$2.7\,Gyr (see Sect.\,\ref{sec:environment}), any lopsided mode would have already dropped \citep{jog09}. Does this imply that secular evolution processes can lead to asymmetries? Since the early works of \cite{bosma78} and \cite{bosmafreeman93} we know that our understanding of a galaxy may change after performing and comparing deep observations that let us reach very low surface brightness (SB or $\mu$) levels of a galaxy and its surroundings. Therefore, this was the natural follow$-$up for CIG\,96. Additionally, as suggested by the N$-$body simulations of \cite{penarrubia05}, the orbital properties of halo substructures are determined by the environment and can survive several gigayears, outliving HI tidal features. Within the last two decades, a number of works have unveiled many faint structures or companions that remained hidden in shallower observations \citep[e.g.][among others]{martinezdelgado08, martinezdelgado09, martinezdelgado15, duc15, vandokkum15, trujillo16, trujillo17, iodice17, bosma17}.


\cite{espada05} also presented the discovery of a close and small companion situated at 15.2$\arcmin$ ($\sim$90 kpc, projected distance) to the east of CIG\,96. To account for the HI asymmetry, they rule out any encounter with a massive companion as well as any close or parabolic passage of another smaller galaxy. They leave the door open for a parallel passage through the equatorial plane of CIG\,96 at an intermediate distance, that is, outside the optical disc but within the extended HI disc. \cite{espada11a} studied the Kennicutt-Schmidt SF law and efficiency in the large atomic HI disc of CIG\,96, using the VLA observations mentioned in this work (see Sect.\,\ref{sec:HIobs}), as well as NUV and FUV observations from GALEX. By comparing the VLA maps and UV images, they found a good spatial correlation between the HI and both NUV and FUV emission, especially outside the inner 1$\arcmin$. Also, the main star-forming regions lie on the enhanced HI emission of two spiral arm-like features that correspond to the HI pseudo-ring. They found that the (atomic) Kennicutt-Schmidt power-law index systematically decreases with the radius. Regarding the star formation efficiency (SFE), they saw that it decreases with radius where the HI component dominates and that there is a break in this correlation at \emph{r} = 1.5$\times\ r_{25}$. However, mostly within the HI pseudo-ring structure, that is, between  1.5$\times\ r_{25}$ and 3.5$\times\ r_{25}$, SFE remains nearly constant. They concluded that this might be a common characteristic in extended UV disc galaxies and that a non-axisymmetric disc can drive the outer spiral arms, as the morphology of the galaxy allows.

In this work we present new and deep HI and optical data of CIG\,96 to study in detail its faint gaseous and stellar components as well as its surroundings in order to reveal any possible causes of its HI asymmetrical distribution and other effects on its evolution. Throughout this study, all mentions to distances between different parts of the galaxy and its surroundings are projected distances unless stated otherwise. Also, we have assumed a cosmology with $H_{0}$ = 75 km\,s$^{-1}$ Mpc$^{-1}$, $\Omega_{\Lambda 0}$ = 0.73 and $\Omega_{m 0}$ = 0.27.


\begingroup
\renewcommand{\arraystretch}{1.2}
\begin{table}
\caption{Parameters of CIG\,96 (NGC 864)}   
\label{table:cig96}
\centering
\begin{tabular}{p{5cm} p{3cm}}        
\hline\hline
\small{Parameter} & \small{Value}\\    
\hline
$\alpha_{(2000)}$\tablefootmark{a} & 02$^{h}$15$^{m}$27.6$^{s}$  \\     
$\delta_{(2000)}$\tablefootmark{a} & +6$^{\circ}$00$\arcmin$09$\arcsec$ \\
Type\tablefootmark{b} & SAB(rs)c \\ 
Distance\tablefootmark{c} & 20.3 Mpc  \\
$r_{25}$\tablefootmark{d} & 2.35$\arcmin$ / 13.9 kpc \\
Inclination\tablefootmark{e} & 46.59$\degree$  \\
$M_{dyn,\ CIG\ 96}$\tablefootmark{e} & 1.78 $\times$ 10$^{11}\ M_{\odot}$ \\
Position angle\tablefootmark{e} & 20.0$\degree$ \\
A$_{int}$($r$)\tablefootmark{c} & 0.185 \\
A$_{k}$($r$)\tablefootmark{c} & 0.006 \\
A$_{int}$($g$)\tablefootmark{c} & 0.255 \\
A$_{k}$($g$)\tablefootmark{c} & 0.008 \\
\hline
\end{tabular}
\tablefoot{\\
\tablefoottext{a}{\cite{leon03}.} \\
\tablefoottext{b}{\cite{deVaucouleurs91}.} \\
\tablefoottext{c}{\cite{fernandezlorenzo12}. Distance computed using $H_{0}$\,=\,75\,km\,s$^{-1}$\,Mpc$^{-1}$. A$_{int}$ and A$_{k}$ represent the internal and k-correction extinction terms in SDSS\,$r$ and $g$ bands.}\\
\tablefoottext{d}{Semi-major axis of the galaxy at the isophotal level 25\,mag\,arcsec$^{2}$ in the $B$\,band \citep{fernandezlorenzo12}.} \\
\tablefoottext{e}{This work. Total dynamical mass $M_{dyn}$ described in Sect.\,\ref{sec:sblimit}.}
}
\end{table}
\endgroup


\section{Description of the observations and data processing}
\label{sec:description}

In this section we present all the HI and optical observations of CIG\,96 used in this work as well as the reduction and calibration processes we followed to obtain the final images. The most relevant data are summarised in Tables \ref{tableHI} and \ref{tableOPT}.

\subsection{HI observations}
\label{sec:HIobs}

In two different epochs, 21 cm line observations of CIG\,96 were made using the NRAO Karl G. Jansky Very Large Array (hereafter VLA or EVLA) observatory.
First, two VLA projects AV0276 and AV0282 were performed in July 2004 and July 2005, respectively. We obtained 3 hours in D-configuration (26 antennas used) and 7 hours in C-configuration (27 antennas used), respectively. Both observing projects had the same set up: 2 IF correlator mode, a bandwidth of 3.125 MHz per IF and a frequency resolution of 48.8 kHz that corresponds to a velocity resolution of 10.4\,km\,s$^{-1}$. 
Second, the Extended-VLA (EVLA) project 13A-341, fully dedicated to observing CIG\,96, was executed during 2013 as follows: 3 hours in March, in D-configuration; 3 hours in May in the hybrid DnC-configuration and 10 hours in July, in C-configuration. In all cases, 27 antennas were used. The set-up of these observations consisted of single IF correlator mode, a bandwidth of 2 MHz and a frequency resolution of 16 kHz, equivalent to a velocity resolution of 3.3\,km\,s$^{-1}$ that was smoothed to 10\,km\,s$^{-1}$ for the calculations. These data are summarised in Table\,\ref{tableHI}.


\begingroup
\renewcommand{\arraystretch}{1.3}
\begin{table*}
\caption{Data and results of the HI observations with VLA/EVLA}
\label{tableHI}      
\centering          
\begin{tabular}{c c c c c c c c}     
\hline\hline     
\small{Telescope/s} &  \small{Velocity range} & \small{Integration time} & \small{Beam size} & \small{Noise} & \small{$N_{HI}$ limit} & \small{$M_{HI}$ limit}\\ 

  &  \small{(km s$^{-1}$)} & \small{($h$)} & \small{($\arcsec \times \arcsec$)} & \small{(mJy beam$^{-1}$, 1$\sigma$)} & \small{(10$^{19}$ cm$^{-2}$, 5$\sigma$)} & \small{(10$^{6}\,M_{\odot}$, 5$\sigma$)} \\ 
\hline
VLA\tablefootmark{a} 				&  1249.5$-$1895.2 & 10 & 27.1$\times$23.6 	 & 0.31 	& 2.68	& 1.5 \\
EVLA\tablefootmark{b} 				&  1330$-$1700 & 13 & 37.6$\times$20.0 	 & 0.84 	& 6.19	& 4.1 \\
VLA$+$EVLA 							&  1330$-$1700 & 19 & 28.0$\times$28.0 		 & 0.25 	& 1.78 	& 1.4 \\
VLA$+$EVLA$+$WF\tablefootmark{c}	&  1330$-$1700 & 19 & 28.0$\times$28.0 		 & 0.13 	& 0.89 	& 0.7 \\
\hline
\end{tabular}
\tablefoot{\\
\tablefoottext{a}{3 h in D configuration, 7 h in C configuration. The original channel width is 10.4 km s$^{-1}$ (48.8 kHz).} \\
\tablefoottext{b}{3 h in D configuration, 10 h in C configuration. The original channel width of 3.3 km s$^{-1}$ was smoothed o 10\,km\,s$^{-1}$ (48 kHz) for the current calibration.} \\
\tablefoottext{c}{Wavelet filtering (WF) applied to the VLA$+$EVLA data.} \\
}
\end{table*}
\endgroup
%


All VLA and EVLA data were fully calibrated and imaged using CASA software package \citep{casa} tasks. We used the CLEAN algorithm \citep{hogbom74} to produce the final datacube. Each data set or measurement set (MS) was scanned to remove bad data and RFI (radio-frequency interferences). They were separately calibrated in phase, amplitude and bandpass and imaged individually to check their suitability for our aims. We produced a set of two individual datacubes by combining all VLA data and all EVLA data, respectively. We discarded the hybrid DnC-configuration data due to the presence of a remarkable amount of RFI, making them too defective for our goals. The VLA data consisted of two individual MSs, one for D-configuration data and one for C-configuration data. EVLA data consisted of thirteen individual MSs: three MSs were obtained in D-configuration and ten MSs in C-configuration. All HI masses in this work have been computed as given by \cite{roberts62} and \cite{roberts75}:

\begin{equation}
\label{eq:mhi}
$$M_{HI}\ (M_{\odot}) = 2.356\times10^{5}\ D^{2}\ S\ \Delta V$$
\end{equation}

where $D$ is distance ($Mpc$) and $S \Delta V$ is the velocity integrated HI flux ($Jy\ km s^{-1}$).

The column density $N_{HI}\ (cm^{-2})$ depends on the brightness temperature $T_{B} (K)$ integrated over the line width $dv$ ($km\ s^{-1}$). In turn, $T_{B}$ depends on the flux density $S$ $(Jy\ beam^{-1})$ and the product of the major and minor axes $Maj \times min$ ($arcsec^{-2}$). Respectively:

\begin{equation}
\label{eq:tb}
$$T_{B}\ (K) = 6.07\times10^{5}\ S \left(Maj\ \times \ min\right)^{-1}$$
\end{equation}

\begin{equation}
\label{eq:nhi}
$$N_{HI}\ (cm^{-2}) = 1.823\times 10^{18} \int T_{B}\ dv$$
\end{equation}

\textit{VLA data cube.}
All VLA data were used to produce a preliminary datacube via imaging using natural weighting. This led to a synthesized beam of 27.11$\arcsec$ $\times$ 23.60$\arcsec$ and a root mean square (rms) noise level of 0.31\,mJy\,beam$^{-1}$ (1$\sigma$), reaching a HI column density limit of $N_{HI} = 2.68\times$10$^{19}$ cm$^{-2}$ (5$\sigma$). Assuming a HI line width of 10 km\,s$^{-1}$, the achieved HI mass detection limit is $\sim 1.5 \times$10$^{6} M_{\odot}$ (5$\sigma$) and a HI column density of 2.7$\times$10$^{19}$ cm$^{-2}$ (5$\sigma$).

\textit{EVLA data cube.}
All EVLA data in C and D configurations were combined and imaged with natural weighting in a preliminary datacube. This datacube had a median rms of 0.84 mJy beam$^{-1}$ (1$\sigma$) and a synthesized beam of 37.57\arcsec $\times$ 19.97\arcsec. Such beam elongation is due to the short right ascension range in which the observations were taken. With a velocity resolution of 10 km\,s$^{-1}$, the HI mass detection limit achieved was of $M_{HI}$ = 4.1$\times$10$^{6}$ M$_{\odot}$ (5$\sigma$).

\textit{Combined EVLA and VLA data cubes: hereafter the HI cube.}
After the rms-weighted\footnote{Weighting computed as $w(i) = rms(i)$ $^{-2}$, where $rms(i)$ stands for the flux density rms of each cube in the same units.} concatenation of the VLA MS and EVLA MS we produced the final datacube of this work (hereafter referred as the HI cube). The corresponding weighting factors applied to the VLA and EVLA data were 10.40 and 1.42, respectively.
The HI cube comprises a total of 19\,hours on target and has a synthesized beam of 28.16\arcsec\,$\times$\,22.72\arcsec\ (2.77\,kpc\,$\times$\,2.24\,kpc at a distance of 20.3\,Mpc); it covers a velocity range from 1330\,km\,s$^{-1}$ to 1800 km\,s$^{-1}$ in 48 channels assuming spectral resolution of 10 km\,s$^{-1}$. We used the kinematical local standard of rest (LSRK) as the frame of reference for the radio velocities. Also, we worked with a smoothed beam of 28\arcsec\,$\times$\,28\arcsec\ to simplify the physical interpretation of the results and avoid beam effects. The corresponding HI cube yielded a median rms of 0.25\,mJy\,beam$^{-1}$\,(1$\sigma$) that allowed us to reach a HI mass limit of $M_{HI}^{lim}\simeq$\,1.4$\times$10$^{6}$\,$M_{\odot}$\,(5$\sigma$) and a HI column density limit of $N_{HI}\simeq$\,1.78$\times$10$^{19}$\,cm$^{-2}$ (5$\sigma$).
After performing a wavelet filtering (see Sect.\,\ref{sec:wavelets}) over the HI cube, we improved these results by a factor of approximately two, reaching a final median rms of 0.126\,mJy beam$^{-1}$\,(1$\sigma$) per channel. The minimum HI mass detected is $M_{HI}^{lim} =$\,0.7$\times$10$^{6}$ $M_{\odot}$\,(5$\sigma$), the HI column density limit is $N_{HI} =$\,8.9$\times$10$^{18}$\,cm$^{-2}$\,(5$\sigma$) and the total HI mass is $M_{HI}^{total} =$\,9.77$\times$10$^{9} M_{\odot}$\,(5$\sigma$). The integrated intensity map, the velocity field and the channel maps are all presented in Sect.\,\ref{sec:HIres}.

\subsection{Wavelet filtering of the HI cube}
\label{sec:wavelets}

A robust detection of faint HI features relies on reaching a column density ($N_{HI}$) that is as low as possible with the best signal$-$to$-$noise ratio (S/N). In order to further improve our $N_{HI}$ limit, we have applied a wavelet filtering to our HI cube which allows to achieve a higher S/N. An in-depth discussion of the wavelet transform is beyond the scope of this paper but we provide here an explanation of the method used in this work. As explained by \cite{leon16}, the wavelet transform is a powerful signal-processing technique that provides a decomposition of the signal into elementary local contributions defined by a scale parameter \citep{grossmann}. The wavelets are the scalar products of shifted and dilated functions of constant shape. The data are unfolded in a space-scale representation that is invariant with respect to dilation of the signal. Such an analysis is particularly suited to studying signals that exhibit space-scale discontinuities and/or hierarchical features, as may be the case for the possible structures located in the outskirts of the HI envelope of CIG\,96.

Following the same procedure as \cite{leon16}, we have used a $B_3$-spline scaling function defined by the following convolution matrix $M$:


\begingroup
\renewcommand*{\arraystretch}{1.8}
\newcommand*\rfrac[2]{{}^{#1}\!/_{#2}}
\[
   M=
  \left[ {\begin{array}{ccccc}
   \rfrac{1}{256} & \rfrac{1}{64} & \rfrac{3}{128} & \rfrac{1}{64} & \rfrac{1}{256} \\
   \rfrac{1}{64}  & \rfrac{1}{16} & \rfrac{3}{32}  & \rfrac{1}{16} & \rfrac{1}{64}\\
   \rfrac{3}{128} & \rfrac{3}{32} & \rfrac{9}{64}  & \rfrac{3}{32} & \rfrac{3}{128}\\
   \rfrac{1}{64}  & \rfrac{1}{16} & \rfrac{3}{32}  & \rfrac{1}{16} & \rfrac{1}{64}\\
   \rfrac{1}{256} & \rfrac{1}{64} & \rfrac{3}{128} & \rfrac{1}{64} & \rfrac{1}{256} \\
  \end{array} } \right]
\]

\endgroup

Similar to the Ricker function (mexican hat), it has a positive kernel surrounded by a negative annulus and the total integrated area is zero.

We have applied this wavelet over the HI calibrated data via the \textit{A trous} algorithm (see \citealt{bij}) as described by \cite{leon00}. This algorithm creates different filtered wavelet planes according to the scale parameters and a certain threshold level. The scale parameters have received values of 2$^{i}$ with $i$\,$\in$\,[1,6], each defining the $i-$th plane. Each $i-$th raw wavelet plane is defined as the subtraction of two components that, in turn, depend on the $i-$th scale parameter: the zeroth component corresponds to the image plane itself; the rest of the $i-$th components are defined as the result of convolving the $i-1-$th component with the previously defined kernel function. The last plane, namely, the last smoothed plane or LSP (in our case, scale parameter of 2$^{6}$) does not undergo any convolution; therefore, it is not a wavelet plane itself but the residuals of the last convolution. With the consequent exception of the LSP, each raw plane is filtered above a threshold to construct the $i-$th filtered wavelet plane. For this work, such a threshold was set at 5$\sigma_{i}$, where $\sigma_{i}$ is the rms noise for the $i-$th plane.

The combination of the filtered wavelet planes and the LSP is possible and may cause the rms to change. Since the original image is spread in different spatial scales, a limited combination of the planes implies the recovered flux will be a lower limit to the total emission contribution. Should all planes be combined, the recovery is complete and the total flux is conserved.

After filtering our HI cube, we combined all planes. The resulting rms and, accordingly, the HI column density limit, were improved by a factor of two, as specified in the last paragraph of Sect.\,\ref{sec:HIobs} and summarised in Table\,\ref{tableHI}.

\subsection{Blanking of the HI cube}
\label{sec:lownoise}

We separated genuine emission from noise by blanking non-signal pixels using the following method.

First, we applied a spatial smoothing over the wavelet-filtered HI cube by convolving it with a Gaussian kernel four times the size of the adopted synthesized beam, that is, 56\arcsec$\times$56\arcsec. The resulting smoothed datacube was only used to create the masks, as described below, and its noise was $rms =$\,0.34\,mJy\,beam$^{-1}$ (1$\sigma$). Second, we created masks for each channel of the smoothed datacube. The shapes of these masks were defined by masking out the pixels with values below a 3.5$\times rms$ threshold ($\sim$1.2\,mJy\,beam$^{-1}$). Finally, the masks from the smoothed datacube were applied to the original datacube (non spatially smoothed) to create the moment maps\footnote{All tasks used to generate the described moment maps are part of the CASA Image Analysis toolkit.}.

This method mainly has two advantageous consequences: one, the depth and spatial resolution of the original datacube remain unaffected by the masking and two, the threshold limit, for the integration, does not take into account the areas in each channel whose only contribution is noise. In other words, the blanking of the HI cube helps us to remove any remaining effect from the side lobes (either positive or negative) that might mimic nonexistent structures.

\subsection{Optical observations}
\label{sec:optobs}
\begin{figure}
\centering
\includegraphics[width=\hsize]{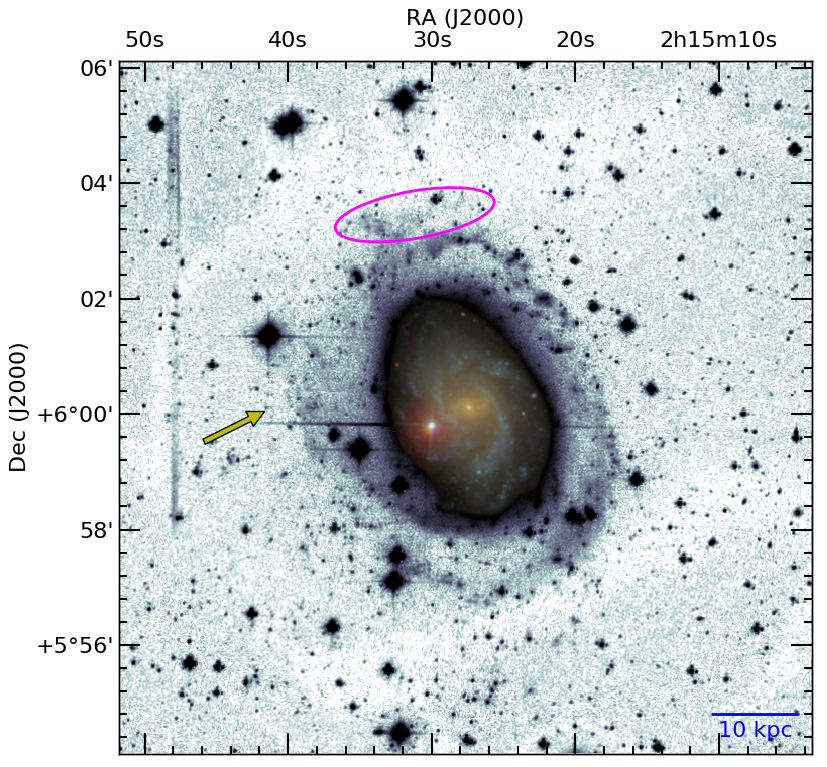}
   \caption{A $12\arcmin \times12\arcmin$ combined image of the Cousins\,$R$ image from CAHA2.2m telescope and the three photographic $B$, $G$, $R$ images from CAHA1.23m telescope. This particular image is only used to show the outer faint structures of the galaxy (e.g. the pseudo-ring, the northern region in a magenta ellipse or the eastern diffuse emission pointed out by the yellow arrow), not for any physical measurement. The inner coloured area corresponds to an SDSS image of CIG\,96 down to $\sim$24\,mag\,arcsec$^{-2}$ (SDSS\,$r$ band) and is used as reference.}
\label{opt22_123}
\end{figure}

\begin{figure}
\centering
\includegraphics[width=\hsize]{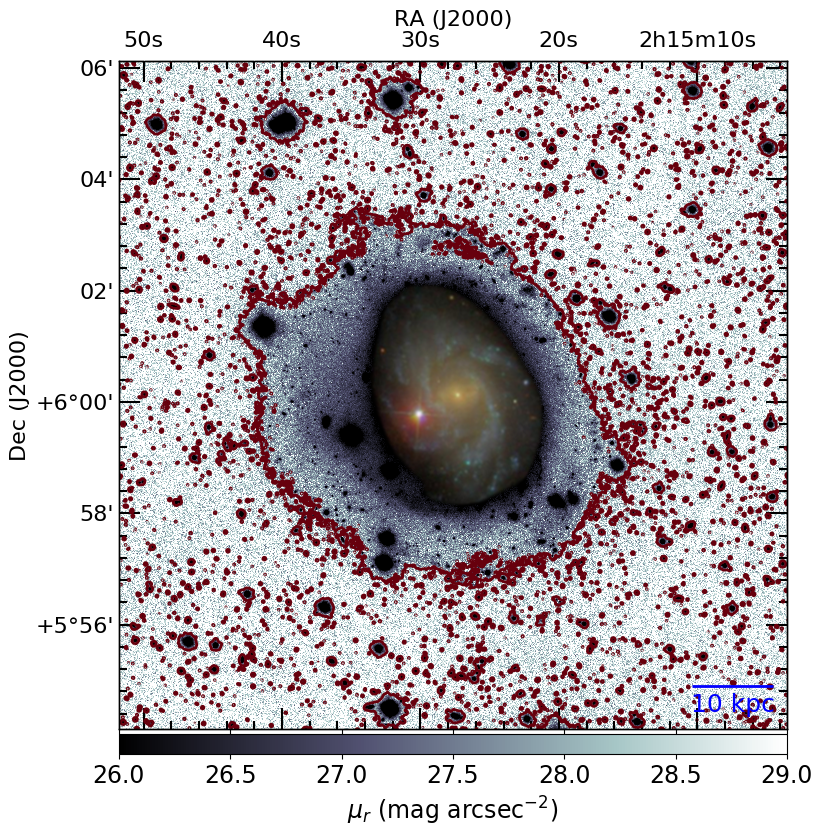}
   \caption{A $12\arcmin \times12\arcmin$ detail of the VST optical image of CIG\,96 with the SDSS colour image down to $\sim$24\,mag\,arcsec$^{-2}$ (SDSS\,$r$). The red contour is set on 26.5\,mag\,arcsec$^{-2}$ (SDSS\,$r$), to point out the faintest SB level of the pseudo-ring.}
      \label{optvst}
\end{figure}

In order to obtain deep optical images of the outskirts and close environment of CIG\,96, we performed observations in three different observatories. Two datasets were observed with the 2.2m and 1.23m telescopes, respectively, at CAHA\footnote{Based on observations collected at the Centro Astron\'omico Hispano Alem\'an (CAHA) at Calar Alto, operated jointly by the Max-Planck Institut für Astronomie and the Instituto de Astrof\'isica de Andaluc\'ia (CSIC).} observatory in Spain. The first dataset is from CAHA2.2m, a deep image with good seeing in the Cousins\,$R$ band (see Sect.\,\ref{sec:22data}). The second dataset consists of three images taken with photographic $B$, $G$, $R$ bands used to study colour index properties (see Sect.\,\ref{sec:123data} and all 2.2m and 1.23m images combined in Fig.\,\ref{opt22_123}). The third dataset was obtained with the VLT Survey Telescope (ESO\footnote{Based on observations made with ESO Telescopes at the La Silla Paranal Observatory under programme ID 093.B-0894 and 098.B-0775.}) in Chile (hereafter, VST) and provides a very deep and wide field image to study the surroundings of the galaxy (see image in Fig.\,\ref{optvst} and Sect.\,\ref{sec:22data}). The most relevant data are summarised in Table\,\ref{tableOPT}.


\begingroup
\renewcommand{\arraystretch}{1.3}
\begin{table*}
\caption{Data of the optical observations}
\label{tableOPT}      
\centering          
\begin{tabular}{c c c c c c c c}     
\hline\hline     
\small{Telescope} & \small{Filter} & \small{Binning} & \small{Spatial scale}\tablefootmark{a} & \small{Total exp. time}\tablefootmark{b}  & \small{Field of view}\tablefootmark{c} & \small{SB limit} & \small{Seeing} \\ 

\small{(Instrument)} & & & \small{($\arcsec$/pixel)} & \small{(\# $\times$ time)} & \small{($\arcmin \times \arcmin$ / kpc$\times$kpc)} & \small{(mag\,arcsec$^{-2}$)} & \small{($\arcsec$)} \\ 
\hline\hline  
CAHA2.2m  & Cousins R & 2$\times$2 & 1.04  & 3h 56m & 12$\times$12 & 27.5 & 1.59 \\
(CAFOS) & & & &  (71 $\times$ 200 s) & 71$\times$71 & & \\
\hline  
CAHA1.23m  & B, G, R & 1$\times$1 & 1.04 & 3h 38m & 15$\times$16 & $-$ & 1.56 \\
(DLR-MKIII) & photographic\tablefootmark{d} & & & (30$_B$, 37$_G$, 42$_R$ $\times$ 120 s)\tablefootmark{e} & 88$\times$94 & & \\
\hline  
VST & SDSS $r$ & 2$\times$2 & 0.21 & 5h 10m & 60$\times$60 & 28.7 & 1.10 \\
(OmegaCAM) & & & & (122 $\times$ 154 s) & 350$\times$350 & & \\
\hline\hline                  
\end{tabular}
\tablefoot{\\
\tablefoottext{a}{Spatial scale according to the binning used.} \\
\tablefoottext{b}{Total number of exposures $\times$ exposure time of each exposure.}\\
\tablefoottext{c}{The top value of each telescope is the field of view in square arcminutes; the bottom value is the field of view according to the distance to the galaxy (see Table\,\ref{table:cig96}).}\\
\tablefoottext{d}{CAHA1.23m images in photographic B and R filters were converted to SDSS\,$g$ and $r$, respectively (see Sect.\,\ref{sec:123data}).} \\
\tablefoottext{e}{The subindex indicates the filter of each corresponding number of exposures. Regardless of the filter, each one has an exposure time of 120 s.} \\

}
\end{table*}
\endgroup



\subsubsection{CAHA2.2m dataset}
\label{sec:22data}
CIG\,96 was first observed in the second half of the night of September 11, 2012, with the CAFOS instrument at CAHA2.2m telescope. The CAFOS SITe1d detector has 2048$\times$2048 pixels with a pixel size of 24\,$\mu m$ (spatial scale of 0.53\arcsec\ \,pix$^{-1}$), providing an effective circular field of view of $\sim$12$\arcmin$ in diameter.

A total of 71 exposures of 200 s each build up a total time on source of 3h 56min. All images were taken in the Cousins\,$R$ filter, dithered by $\sim$20\arcsec\ and in 2$\times$2 binning mode, providing a pixel scale of 1.04\arcsec /pixel. The night conditions were photometric during most of the night, with a median seeing of 1.59\arcsec\ (seeing ranging from 1.31\arcsec\ to 1.81\arcsec). We used standard reduction and calibration techniques from \textit{repipy} and \textit{LEMON} packages \footnote{\textit{repipy} (\href{https://github.com/javierblasco/repipy}{GitHub source}) reduction package by J. Blasco-Herrera, \textit{LEMON} (\href{https://github.com/vterron/lemon}{GitHub source}) calibration package by V. Terr\'on-Salas.} and IRAF.
No standard stars were measured in this campaign and so the extinction coefficient was computed by means of non-saturated stars present within the field of view of our observations. As a consequence, a larger uncertainty is introduced in the photometric calibration. In order to obtain the Zero Point of the night, we computed the Bouguer fit of eight non-saturated stars (visible in all images) and calibrated them with the corresponding data from SDSS \citep{ahn12}. Since this dataset was taken using Cousins\,$R$ filter, all fluxes were converted from SDSS magnitudes system to Cousins\,$R$ using the transformation by \cite{lupton}, derived by matching photometry data from SDSS Data Release 4 (DR4) to Peter Stetson's published photometry for stars:
\begin{equation}
\label{eq:sdssRri}
\begin{gathered}
$$R_{ri} = r - 0.2936*(r-i) - 0.1439$$
\end{gathered}
\end{equation}

in magnitudes, where $r$ and $i$ are the magnitudes in the SDSS\,$r$ and SDSS\,$i$ filters, respectively. The median Zero Point of the night (Cousins\,$R$ filter) is 24.28\,$\pm$\,0.12\,mag. 
We calculated the SB of the image by setting 40 square boxes of 20\arcsec$\times$20\arcsec size in the southern, western and northern areas of the image. The eastern side of the CAHA2.2m image is heavily contaminated by a star so we did not take into account any SB measurements of that side. There is a slightly uneven distribution of the light between the western side (median $\mu_{Cous\ R}=$\,27.5\,mag\,arcsec$^{-2}$) and the northern and southern sides ($\mu_{Cous\ R} = $\,28\,mag\,arcsec$^{-2}$). We cannot confirm whether the 0.5\,mag\,arcsec$^{-2}$ difference comes from the residuals of the flat-fielding or from reflected light and the small field of view of the image prevents selecting a SB value over the rest so we set the SB limit of the image as the lowest value, $\mu_{Cous\ R}=$\,27.5\,mag\,arcsec$^{-2}$ (approximately $\mu_{SDSS\ r} =$\,28.0\,mag\,arcsec$^{-2}$).


\subsubsection{CAHA1.23m dataset}
\label{sec:123data}
CIG\,96 was observed for a second time on the night of December 8, 2012 with the DLR-MKIII instrument at the CAHA 1.23m telescope. The camera is equipped with an e2v CCD231-84-NIMO-BI-DD sensor (4k$\times$4k pixels, 15\,$\mu m$\,pix$^{-1}$). The original field of view is 21.5\arcmin$\times$21.5\arcmin\ but the observations were cropped down to the central 15\arcmin$\times$16\arcmin.

In this case, we used three different filters: photographic B, G and R (different from Johnson-Cousins) for which a total of 30, 37 and 42 exposures of 120\,s each were taken, respectively, in 1$\times$1 binning mode. The night conditions were stable for the most part of the night and all filters present a median seeing of 1.56\arcsec\ (seeing range from 1.48$\arcsec$ to 1.61$\arcsec$). The total integration time was 3h\,38min. 


As with the previous dataset, standard reduction was applied to all the images in each filter separately. However, they were divided by a blank field. It was obtained from an adjacent galaxy-free field and corrected for bias and regular flat field too, so the remaining image would not show any residual gradient. Dividing the images by this blank field allows large-scale structures to be removed. 
We used the SDSS tabulated fluxes from several stars to calibrate the images via the following relation between SDSS and photographic filters: B(3900-5100 angstrom) would correspond to SDSS\,$g$ and R(5800-7000 angstrom) to SDSS\,$r$. However, G(4900-5800 angstrom) would lie right between SDSS\,$g$ and $r$ bands. 
For the conversion of G band to SDSS, we considered different scenarios in which the emission was split between SDSS\,$g$ and $r$ bands but it has not been used further in this work. 
Hereafter we focus on the empirical relations that we calculated for R and B bands with respect to SDSS\,$r$ and $g$. The initial relations between the corresponding magnitudes (not corrected from extinction) are:
\begin{equation}
\label{eq:rRnoext}
$$m^{+ext}_{r_{\ SDSS}} = 1.01*m_{R_{phot}} - 9.83 \pm0.15$$
\end{equation}
and
\begin{equation}
\label{eq:gBnoext}
$$m^{+ext}_{g_{\ SDSS}} = 0.99*m_{B_{phot}} - 9.70 \pm0.33$$
\end{equation}

Internal extinction and $k-$correction were applied to the fluxes in both $g$ and $r$ bands. We used the extinction laws by \cite{savage79} (in agreement with \citealt{fitzpatrick99}) where $A(B) = 4.10 \times E_{B-V}$; the internal extinction and $k$-correction in the B band for CIG\,96 are $A_{int}(B) = 0.276$ and $A_k(B) = 0.009$, respectively \citep{fernandezlorenzo12}; the extinction-reddening relations for the SDSS bands are $A_x(g) = 3.793 \times E_{B-V}$ and $A_x(r) = 2.751 \times E_{B-V}$ \citep{stoughton02}.

These relations yield the following internal and $k$-correction values for each band: $A_{int}\,(g)\,=\,0.255$, $A_{int}(r)\,=\,0.185$, $A_k(g)\,=\,0.008$ and $A_k(r)\,=\,0.006$.

Hence, the final empirical extinction$-$corrected equations that convert photographic B and R bands to SDSS\,$g$ and SDSS\,$r$ bands are:
\begin{equation}
\label{eq:rR}
$$m_{r_{SDSS}} = 1.01*m_{R_{phot}} - 10.02 \pm0.15$$
\end{equation}
and:
\begin{equation}
\label{eq:gB}
$$m_{g_{SDSS}} = 0.99*m_{B_{phot}} - 9.96 \pm0.33$$\\
\end{equation}

Finally, the images were average stacked applying an outlier-rejection algorithm.

With the two images from B and R bands already calibrated to SDSS\,$g$ and SDSS\,$r$ bands respectively, we built a $g-r$ image with the aim of studying the colour distribution in the most interesting regions of the galaxy (see Sect.\,\ref{sec:pseudocolor}).

In Fig.\,\ref{opt22_123} we show the result of combining the reduced CAHA2.2m image (Cousins\,$R$ band) and the three reduced CAHA1.23m images (photographic filters). The lower resolution of these images (compared to the better resolution of VST, see Sect.\,\ref{sec:vstdata}) provides a more clear visualization of the external structures of CIG\,96, especially the faint structure in the N and the very diffuse E side of the pseudo-ring, indicated in the image. However, we cannot calibrate them all to a common band, so this image must be taken only as an illustrative view of the galaxy.

\subsubsection{VST dataset}
\label{sec:vstdata}
In order to study the larger-scale structure surrounding CIG\,96, we also observed the galaxy with OmegaCAM at the VST (runID: 098.B-0775(A)).
This instrument has a field of view of 1 square degree sampled with a 32-CCD, 16k$\times$16k detector mosaic at 0.21\arcsec/pix.

The 32 CCDs have intermediate spaces between the different chips in the vertical direction (5.64\,mm top and bottom gaps; 0.82\,mm central gap) and in the horizontal direction (1.5\,mm gap). Also, at the time these observations were designed, the user manual accounted for cross talk between CCDs 93 to 96 at <\,0.4\% level (slightly above our aim of 0.35\%). Further discussion with the telescope staff alerted to irregular gain variations in CCDs 82, 87 and 88. In order to avoid these CCDs as much as possible and guarantee a homogeneous coverage of the gaps, we initially designed a manual diagonal dithering pattern for the pilot observations exposures to sample the galaxy and its surroundings. With it, the 49 different offset positions of the galaxy (7 pointings with 7 offset positions each) were placed along a diagonal oriented from the southeast (SE) to the northwest (NW) of the chip, always leaving at least 1\arcmin (both in RA and Dec) with the edge of the CCDs. After the pilot observations, we concluded that the previous diagonal dithering would not significantly differ from the already existing modes (JITTER and DITHER, since STARE was not useful for our aims) so we designed a new manual dithering pattern that would make a total of seven pointings, six of them to the corresponding apexes of a slightly irregular hexagon-shaped pattern plus one more central pointing.


A total of eight observing blocks (OBs) of 1 hour each were dedicated to observing CIG\,96. From these, 7\,OBs had 16 exposures and 1\,OB had 10 exposures, making a total of 122 exposures of 154\,s each. 
The 8\,OBs were carried out on the nights of October 6, 9 and 20, November 1 and 2 and December 2, 3 and 20, all in 2016. The total time spent on source was 5h 10 min. All observations were done under the following conditions: photometric sky transparency, maximum seeing of 2.0\arcsec, airmasses below 2.0 (>63.4$\degree$), with an angular distance to the Moon of at least 60$\degree$ and its maximum illumination at 30\%. We used a modified version \citep{venhola17} of Astro-WISE pipeline \citep{mcfarland13} to reduce and calibrate these data. The SB of the image was calculated as the median of the SB values computed in $\sim$60 square boxes of 20\arcsec$\times$20\arcsec spread in the central 40\arcmin$\times$40\arcmin of the image and avoiding stars. For this we used equation \ref{eq:SBvst}, in which the second term corresponds to the conversion from pix$^{-2}$ to arcsec$^{-2}$:

\begin{equation}
\begin{split}
    \label{eq:SBvst}
        \mu_{SDSS\ r} &= -2.5*log(F_{SDSS\ r}) - 2.5*log(0.21^{2}) \\ \\
                      &= -2.5*log(F_{SDSS\ r}) + 3.3889
\end{split}
\end{equation}

Figure\,\ref{optvst} shows an SDSS colour image of CIG\,96 on top of a subset of the VST image. Additionally, the faint SB reached with this image allowed us to detect Galactic cirrus around the galaxy (see Sect.\,\ref{sec:cirrus}).


\begin{figure}
\centering
\includegraphics[width=\hsize]{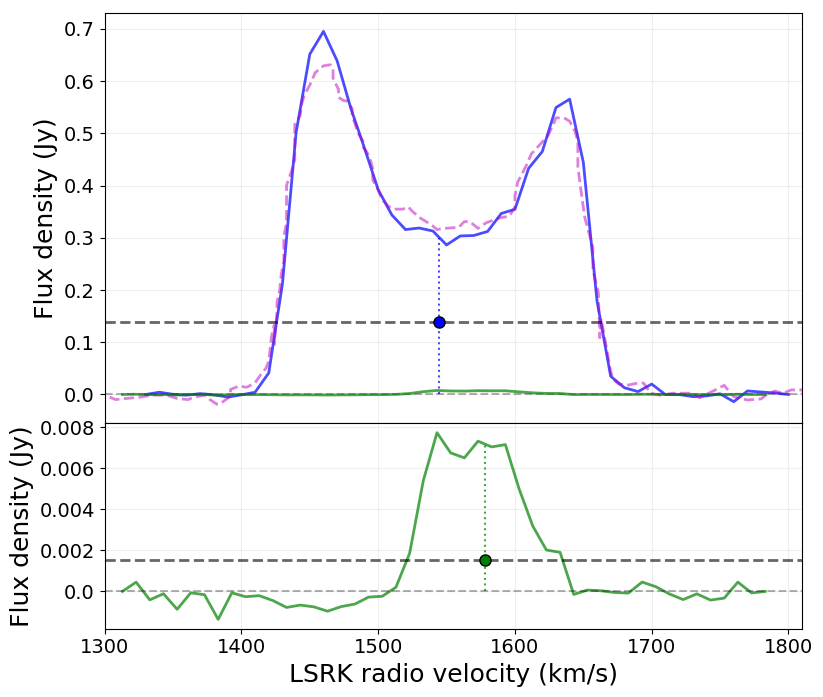}
   \caption{\textit{Top panel}: Integrated profile of CIG\,96, calculated from the EVLA and VLA combined HI cube (blue solid line), integrated spectrum of CIG\,96 (LSRK) obtained by \cite{haynes98} at Green Bank 43 m (heliocentric) (pink dashed line). Our integrated spectrum shows a central velocity that is lower than the Green Bank spectrum, therefore in order to match and facilitate the comparison between the two, we have shifted the latter by $-$17\,km\,s$^{-1}$. The green solid line is the integrated profile of the closest companion of CIG\,96: NGC\,864\,COM01. The horizontal green dashed line sets the width at 20\% of the highest flux peak ($W20$) for the central radio velocity computation, shown as a blue dot ($V_{LSRK}\ (CIG\ 96) = 1544.15$ km\,s$^{-1}$). The vertical blue dotted line defines the two halves of the spectrum for the asymmetry parameter calculation. \textit{Bottom panel}: Integrated HI profile of the companion NGC\,864\,COM01 with a rescaled flux density for an easier visualization. The green dot sets the central velocity of this galaxy ($V_{LSRK}\ (companion)$ = 1577.90 km\,s$^{-1}$).}
      \label{lsrk}
\end{figure}


\subsection{Planck and WISE images}
\label{sec:planckwise}

In order to inspect the cirrus around CIG\,96 (see Sect.\,\ref{sec:cirrus}), we used images from the HFI camera of the Planck satellite at 857\,GHz / 350\,$\mu m$ band \citep{planck14}. Also, we have used a WISE band 3 image (12\,$\mu m$) since this band that traces hot dust and shows good correlation with the cirrus emission \citep{miville16}. Throughout this work, we will refer to these images as Planck857 and WISE3, respectively. Planck857 images were obtained from SkyView online tool \citep{mcglynn94} while the WISE3 image was obtained from the IRSA, NASA/IPAC archive and was reprocessed to improve the flat fielding and remove the stars.


\begin{figure*}
   \resizebox{\hsize}{!}
            {\includegraphics[scale=0.5]{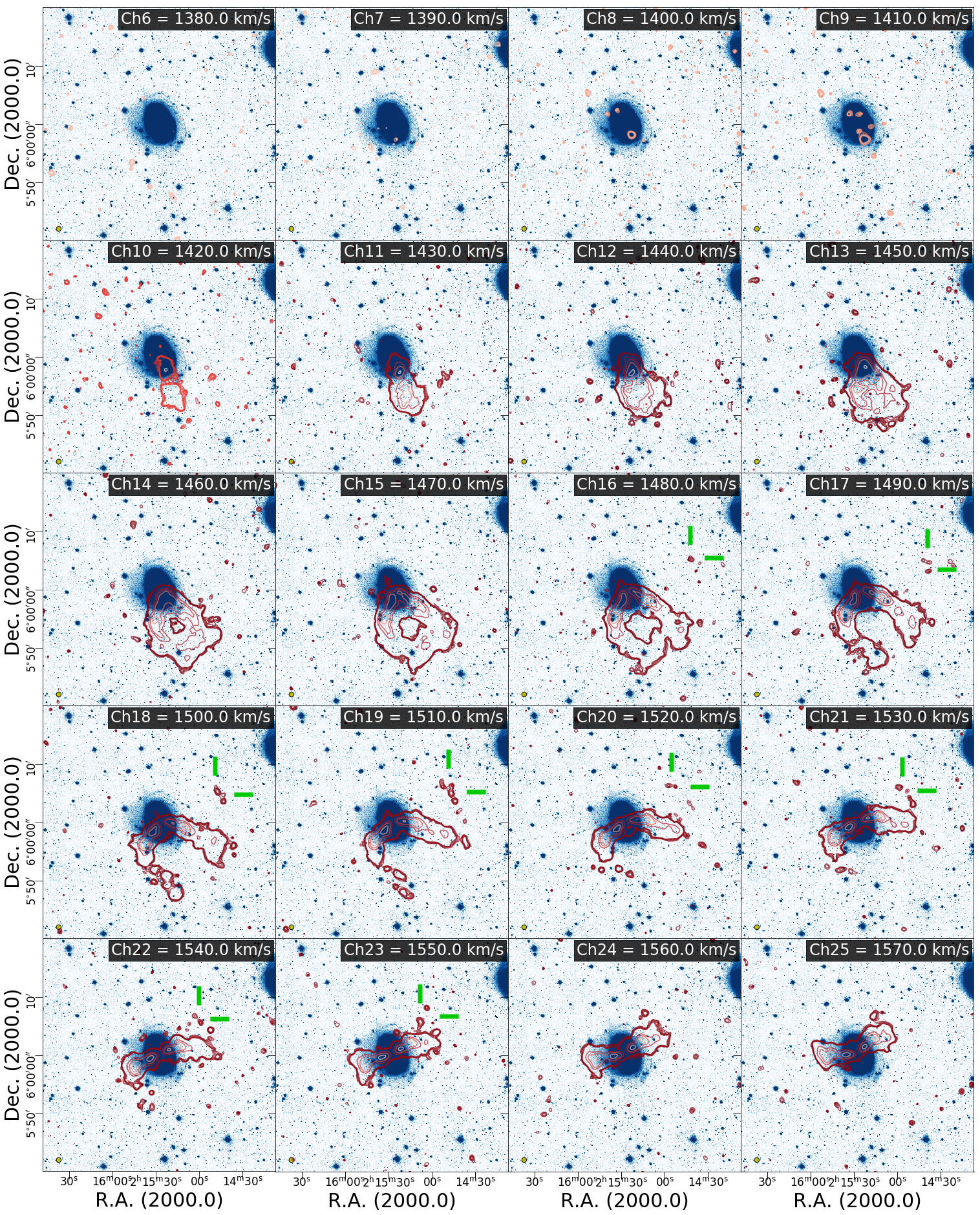}}
      \caption{Channel maps of the wavelet filtered HI cube superimposed on the VST optical image of CIG\,96. The field of view is approximately 25$\arcmin$ $\times$ 25$\arcmin$ (147$\times$147 kpc). \textit{Foreground}: red contours correspond to 3.4, 3.9, 4.5, 5.1, 5.6, 28.1, 56.2, 112.5 and 224.9 $\sigma$ levels (rms = 0.126 mJy beam$^{-1}$, 1$\sigma$) or the equivalent HI column densities of 0.6, 0.7, 0.8, 0.9, 1.0, 5.0, 10.0, 20.0, 40.0 $\times$10$^{19}$ cm$^{-2}$, respectively. Green and magenta marks indicate the NW and SE HI features, respectively. The synthesized beam of 28$\arcsec$ $\times$ 28$\arcsec$ is shown in the bottom left corner as a yellow circle. \textit{Background}: VST image of CIG\,96. We display a SB range of $\mu_{r\ SDSS} = $ 26.0 $-$ 28.4\,mag\,arcsec$^{-2}$ to enhance the outskirts of the galaxy while brighter inner structures are not shown.}
         \label{figvelchannelwz1}
\end{figure*}
   
\begin{figure*}
  \ContinuedFloat
   \resizebox{\hsize}{!}
            {\includegraphics[scale=0.5]{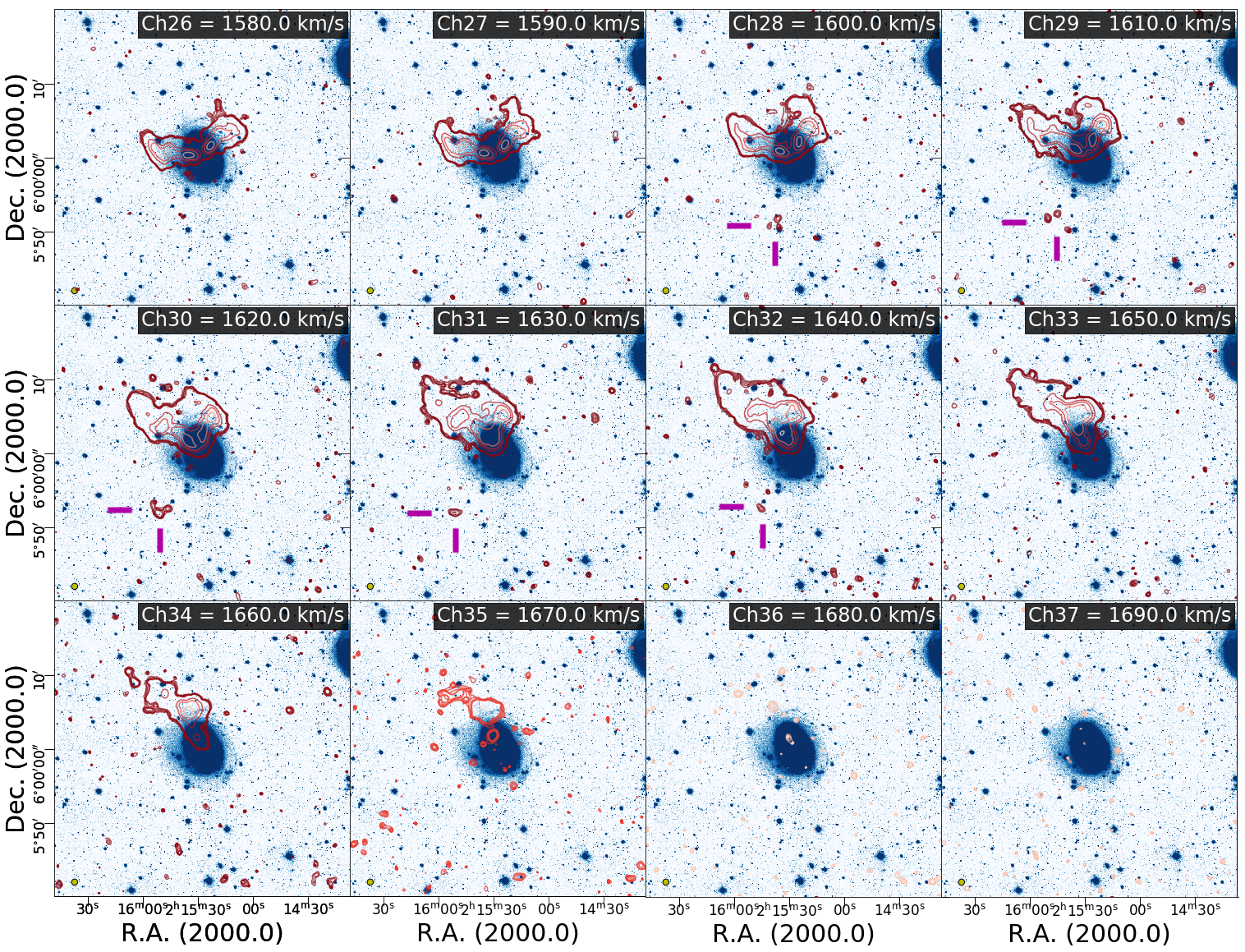}}
      \caption{\emph{Continued}.}
         \label{figvelchannelwz2}
\end{figure*}


\section{HI results}
\label{sec:HIres}

\subsection{Integrated emission and asymmetry level}
\label{sec:velasym}

To calculate the total spectrum, we integrated the emission of each channel of the HI cube. Then, as discussed by \cite{fouque90}, we computed the central velocity of the galaxy $V_{cen}$ as the average between the lowest and highest velocities measured at a width (or flux level) of the 20\% of the highest flux peak in the integrated spectrum (abbreviated $W20$, name varies depending on the percentage used). The error can be estimated as:
$\Delta V$\,=\,4\,$\frac{\sqrt{\delta\nu\,(W20 - W50)/2}}{S/N_{peak}}$ where $\delta\nu$ is the spectral resolution of the cube, $(W20 - W50) / 2$ represents the steepness of the edges of the HI profile at 20\% and 50\% of the maximum flux, and $S/N_{peak}$ is the S/N of the maximum flux peak. Taking these into account, the $W20$ central radio velocity of our HI cube is $V_{LSRK} (CIG\,96)$ = 1544.15 $\pm$ 0.23 km\,s$^{-1}$.

We find a difference of approximately 10 km\,s$^{-1}$ between our result for the central velocity of CIG\,96 and those calculated from single-dish data by \cite{espada05} (same method as in this work) and \cite{haynes98}, 1561.6 and 1562 $\pm$ 1 km\,s$^{-1}$, respectively, both in heliocentric frame of reference, that is, approximately 1553 and 1554 km\,s$^{-1}$ when converted to LSRK, as is ours. \cite{kerr86} also provide a LSRK velocity of 1553 $\pm$ 1 km\,s$^{-1}$, showing  the same shift with respect to our result. To identify the reason for this apparent inconsistency, we recalculated the central velocity of our HI cube and the one published by \cite{espada05} in different standards of rest and in the two optical and radio velocity conventions. In all cases, the differences remained within a few km\,s$^{-1}$, i.e., no change in the standard of rest or velocity convention would account for such a shift. The calibration process was also revised and the correct rest frequency for the HI line was confirmed, leaving us with the only hypothesis of an undetected error in the raw data or the calibration process. Taking this into account, we conclude this difference may be assumed, not to affect the interpretation of the data in any case since it is a small shift compared to the width of the profile.

CIG 96 has a close companion: NGC\,864\,COM01 (hereafter also referred to as the companion), detected in HI by \cite{espada05}. We determine a $W20$ central radio velocity of $V_{LSRK}$ (companion) = 1577.90 $\pm$ 2.62 km\,s$^{-1}$. The HI and optical properties of this galaxy, as well as its implication in the isolation of CIG\,96, are discussed further in Sect.\,\ref{sec:cigcomp}, \ref{sec:companion} and \ref{sec:environment}.

In Fig.\,\ref{lsrk} we compare the integrated emission spectra derived from our HI cube for CIG\,96 and its companion with the one obtained by \cite{haynes98} using data from Green Bank 43m single dish telescope, and still in the heliocentric system of reference. For a better comparison between the two spectra, we have shifted the latter by $-$17 km\,s$^{-1}$. The perfect match between them strongly suggests that our HI cube has a velocity shift of $-$10 km\,s$^{-1}$, after converting all velocities to LSRK.

In order to estimate the HI asymmetry level of a galaxy, quantified as A$_{flux\ ratio}$ \citep[e.g.][]{haynes98, kornreich01, espada11b} we also use the HI integrated spectrum. A$_{flux\ ratio}$ is an areal asymmetry parameter defined as the emission ratio set between the two regions of the spectrum defined with respect to the central velocity and its lowest and highest velocity. While it provides a simple quantification of the gas distribution in the two halves of the galaxy, this global parameter does not give spatial information of any possible asymmetry.


We calculated the sources of the uncertainties of this parameter as described by \cite{espada11b}, obtaining A$_{flux\ ratio}$\,=\,1.16\,$\pm$\,0.01, that is, 16 $\pm$ 1\%, in full concordance with \cite{espada05}.


\subsection{Channel maps}
\label{sec:chmap}

The channel maps allow to inspect every channel of the HI data cube. Each one corresponds to a different velocity allowing us to trace any structures that might be connected to the gaseous envelope of the galaxy. In Fig.\,\ref{figvelchannelwz1} we show a subset of the channel maps of the wavelet filtered HI cube on top of the CAHA1.23m optical image (band R) of CIG\,96. This image corresponds to the central $25\arcmin \times 25\arcmin$ of the primary beam and to the channels with emission, that is, from 1380 to 1690 km\,s$^{-1}$ (channels 6 to 37, respectively) where the channel width is 10 km\,s$^{-1}$. The systemic velocity of the galaxy ($V_{LSRK}\ (CIG\ 96) = 1544.15$ km\,s$^{-1}$, see Sect.\,\ref{sec:velasym}) corresponds to channel 23 and the approaching and receding sides of the galaxy extend approximately 135 and 145 km\,s$^{-1}$, respectively. The synthesized beam is 28$\arcsec \times28\arcsec$, the rms is 0.126 $mJy\ beam^{-1}$ and the column density reached is $N_{HI} (5\sigma) = 8.9\times10^{18} cm^{-2}$.

The HI distribution is more symmetrical in the central channels ($\sim$1500 $-$ 1600 km\,s$^{-1}$) than in those with velocity differences of $\Delta V \geqslant$\,60 km\,s$^{-1}$ with respect to the central velocity. In the latter, the approaching side shows that the HI has a uniform distribution over a larger area in the southwest (SW) than in the receding side, where the distribution is more narrow and oriented towards the northeast (NE). The HI extension also differs, reaching $\sim$7.9\arcmin ($\sim$47 kpc) towards the SW and $\sim$9.3\arcmin ($\sim$55 kpc) towards the NE. Also, the receding NE side is less massive, as reflected in the asymmetrical shape of the integrated spectrum (Sec \ref{sec:velasym}). In both the approaching and receding sides, the HI is extended beyond 4$\times r_{25}$ of the optical extension. From 1630 to 1670 km\,s$^{-1}$ (channels 31 to 35), there is a change in the orientation of the HI, especially visible in column densities below 1.0$\times10^{19}$ cm$^{-2}$ (outer contours of Fig.\,\ref{figvelchannelwz1} and moment maps shown in Sect.\,\ref{sec:HImoments}).

Focusing on the outermost regions, we note two previously undetected features:
\begin{itemize}
\item[$\bullet$] First, from 1480 to 1550 km\,s$^{-1}$ (channels 16 to 23), we notice a clumpy structure to the NW of the galaxy ($\alpha = 02^{h}15^{m}05.9^{s},\ \delta = 6\degree03\arcmin03\arcsec$), with an approximate size of $\sim$21 kpc ($\sim$3.5 arcmin, measured from channel 17 to 22), a column density of approximately $N_{HI}^{NW} \simeq$ 6.5$\times$10$^{19}$ cm$^{-2}$ and a total HI mass of $M_{HI}^{NW\ feat.} \simeq 3.1 \times 10^{6} M_{\odot}$. We refer to this as the NW HI feature and it is indicated with green marks in Fig.\,\ref{figvelchannelwz1}.

\item[$\bullet$] Second, from 1600 to 1640 km\,s$^{-1}$ (channels 28 to 32), a structure shows up to the SE of the galaxy ($\alpha = 02^{h}15^{m}41.0^{s},\ \delta = 5\degree55\arcmin31\arcsec$), within a square region of approximately 8.8$\times$8.8 kpc ($\sim$90\arcsec$\times$90\arcsec) size, a column density of approximately $N_{HI}^{NW} \simeq$ 4.9$\times$10$^{19}$ cm$^{-2}$ and a total HI mass of $M_{HI}^{SE\ feat.} \simeq 1.6 \times 10^{6} M_{\odot}$. We refer to this as the SE HI feature and it is indicated with magenta marks in Fig.\,\ref{figvelchannelwz1}.
\end{itemize}
These structures are discussed further in Sect.\,\ref{sec:feats96}.

\subsection{NGC\,864\,COM01, the HI rich companion of CIG\,96}
\label{sec:cigcomp}

As described by \cite{espada05} and introduced in Sect.\,\ref{sec:velasym}, CIG\,96 has a small companion located at 15.2$\arcmin$ ($\sim$90 kpc) to the east with a \textit{B} magnitude of $m_{B}=$ 16.38 mag. It shows emission throughout 11 channels (from 1540 to 1650\,km\,s$^{-1}$). Its central LSRK velocity is of $V_{LSRK}=$ 1577.90 km\,s$^{-1}$ and a total HI mass of $M_{HI} = 5.1\times10^{6}\ M_{\odot}$. The HI image of this galaxy is shown in Figs.\,\ref{mom0} and \ref{mom1}.
Both CIG\,96 and its companion share a similar orientation of their minor axis. However, they show different kinematical orientation, that is, the companion is counter-rotating with respect to CIG\,96, and we do not find any signs of tidal features between them. The galaxy is studied further in Sect.\,\ref{sec:companion}.

\subsection{Moment maps and position-velocity profiles}
\label{sec:HImoments}

\begin{figure}
\centering
\includegraphics[width=\hsize]{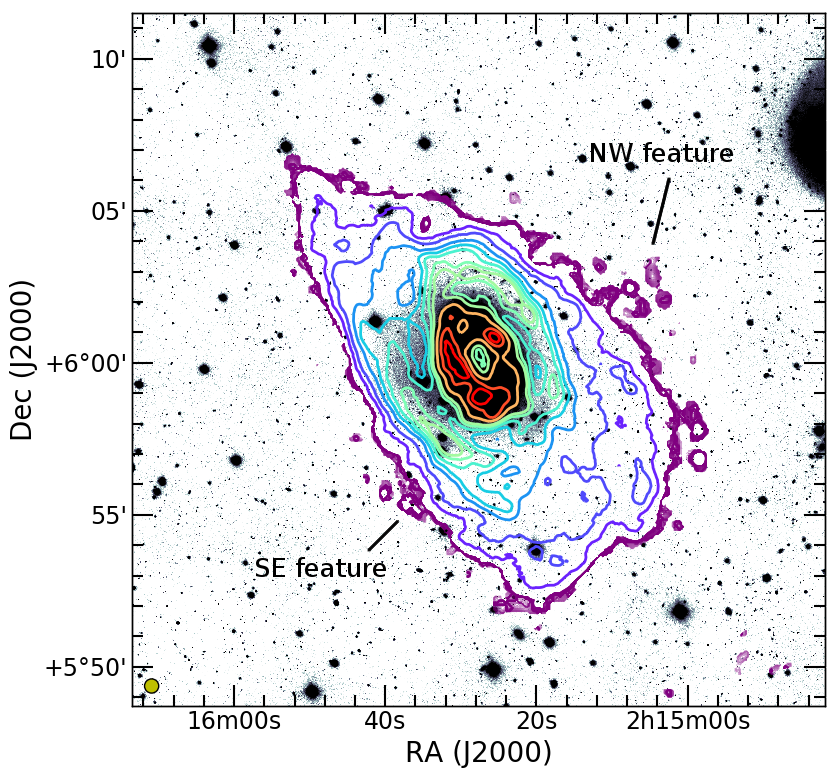}
   \caption{\textit{Background}: VST optical image of CIG\,96 ranging from 26 to 28\,mag\,arcsec$^{-2}$. \textit{Foreground}: HI cube integrated profile contours showing column densities of 0.6, 7.1, 14.1, 28.2, 42.3, 56.5, 70.6, 80.4, 105.8, 127.0 and 141.1 $\times10^{20}$ cm$^{-2}$. The yellow circle at the bottom left represents the beam size of 28\arcsec $\times$ 28\arcsec.}
      \label{mom0opt}
\end{figure}

The integration of the flux density $S$ (or zeroth moment) is carried out from channel 6 (1380 km\,s$^{-1}$) to channel 38 (1700 km\,s$^{-1}$), i.e. one additional channel beyond the HI emission. The velocity field (or first moment) is the intensity-weighted velocity of the spectral line, i.e., a measure for the mean velocity of the gas.
The zeroth moment is shown in Figs. \ref{mom0} and \ref{mom0opt}. The HI extends beyond 4$\times r_{25}$, that is, approximately up to 50 kpc (8.5$\arcmin$), reaching an integrated column density of $N_{HI}\ (5\sigma) = 1.2\times10^{20}$ cm$^{-2}$ with a beam size of 28\arcsec $\times$ 28\arcsec. As a comparison, in Fig.\,\ref{mom0} we indicate with a black line the approximate $N_{HI}\ (5\sigma) = 8.7\times10^{20}$ cm$^{-2}$ column density reached by \cite{espada11a} with a beam size of 16.9\arcsec $\times$ 15.6\arcsec. Quantitatively, the current observations are roughly seven times deeper than the previous ones.

The first moment is shown in Fig.\,\ref{mom1}. It allows the estimation of the position angle (from now on, PA) of the major and minor kinematical axes of the galaxy, indicated by the two black lines at PA\,=\,20$\degree$ and PA\,=\,110$\degree$, respectively.

We have performed the position-velocity (P/V) profiles over the HI cube along the major and minor axes, as shown in Fig.\,\ref{pvcuts}. The emission located at the largest radius in the SW region (indicated with a cyan arrow in the profile over the major axis, Fig.\,\ref{pvcuts}, top panel) was already detected by \cite{espada05}. It is visible in the channel maps at 1450 - 1470 km\,s$^{-1}$ (channels 13 to 15) and it shows a drop in velocity of about 30 $-$ 40 km\,s$^{-1}$ with respect to inner parts of the galaxy. Both its extension and velocity drop are in agreement with the previous work. The interruption in the emission to the NE is due to a $\sim$3$\times$3 kpc$^{2}$ region ($\sim$30\arcsec$\times$30\arcsec) with low HI emission. It is visible in the zeroth moment map ($RA = 2^{h}15^{m}34.935^{s}, DEC = 6\degree04\arcmin33.17\arcsec$) as well as in the channel maps at 1630 - 1640 km\,s$^{-1}$ (channels 31 and 32).

The P/V profile over the minor axis cuts through part of the NW HI feature (indicated with a red arrow, Fig.\,\ref{pvcuts}, bottom panel), the clumpy HI structure mentioned in Sect.\,\ref{sec:chmap}. This feature shows a velocity gradient of $\sim$70 km\,s$^{-1}$ (approximately from 1480 to 1550 km\,s$^{-1}$) and it seems to connect with the galaxy in the channels around its central velocity (channels 23 to 25). Also, the central part of the galaxy shows emission in a wide range of velocities with respect to the central velocity. We discuss this effect further in Sect.\,\ref{sec:feats96}. 


\section{Optical data results}
\label{sec:Optres}

\subsection{Surface brightness limit, dynamical masses and optical features}
\label{sec:sblimit}

The images from CAHA2.2m and CAHA1.23m telescopes have a field of view of 12$\arcmin \times$12$\arcmin$, i.e. approximately 71$\times$71 kpc (see Fig.\,\ref{opt22_123}), while the VST covers 1$\degree \times$1$\degree$, that is, approximately a 350$\times$350 kpc field centred on the galaxy. The limiting SB reached is deeper than any other previously published, in particular with the VST image ($\mu_{r\ SDSS}$\,$(VST)$\,=\,28.7\,mag\,arcsec$^{-2}$, see Fig.\,\ref{optvst}) that reveals unprecedented detail of the extension, boundaries and structures of the external and faint pseudo-ring of CIG\,96 as well as its connection to the inner parts of the galaxy. The VST image also shows signs of Galactic cirrus (See Sect.\,\ref{sec:cirrus}) so we set our reliable detection limit in $\mu_{r\ SDSS}$\,$(VST)$\,=\,28.4\,mag\,arcsec$^{-2}$, just above the level where they start to become visible.

The total dynamical mass of CIG\,96 is $M_{dyn,\ CIG\ 96}$\,=\,1.78\,$\times$10$^{11}$\,M$_{\odot}$, following the calculation described by \cite{cou14}. It was estimated taking into account the inclination ($i$, in degrees, indicated in Table\,\ref{table:cig96}), the radius of the galaxy along the major axis ($R$, in kpc) as well as the rotation velocity ($V$, in km\,s$^{-1}$). Both $R$ and $V$ are extracted from the HI data: $R$ of 6\arcmin\ (35.43\,kpc) from the rotation curve of the major axis (see Sect.\,\ref{sec:HImoments}) and $V$ via measuring the velocity difference at such radius with respect the central velocity of the galaxy, resulting in 125\,km\,s$^{-1}$. 
The same calculation was made for the companion. We obtained a P/V cut of the galaxy along a PA of 35\degree\ to measure the peak $R$ and $V$, resulting in 35\arcsec\ (3.44\,kpc). However, with the current data we do not observe a turn over in the rotation curve so the mass calculation at this radius must be taken as a lower limit. We also assumed an inclination of 90\degree\ since it might be an edge-on galaxy (discussed further in  Sect.\,\ref{sec:companion}). The velocity extent measured at a 35\arcsec\ radius is of 60\,km\,s$^{-1}$. The dynamical mass of the companion is of $M_{dyn,\ comp}$\,=\,2.88\,$\times$10$^{9}$\,M$_{\odot}$. Hence, the dynamical mass relation between the host galaxy and its companion is approximately of $M_{dyn,\ CIG\ 96}/M_{dyn,\ comp}\simeq$\,62. The case of CIG\,96 can be considered similar to the one of the MW-mass galaxy M94 that, after a deep search performed as part of the recent work by \cite{smercina18}, only shows two satellites.

The brightest stellar structures within the pseudo-ring ($\mu_{Cous\ R}\ (CAHA)$ = 25.5 $-$ 26.5\,mag\,arcsec$^{-2}$) are located within a distance of \emph{r}\,=\,1.5\,$-$\,2.0\,$\times$ $r_{25}$ from the galaxy centre (i.e. approximately 3.5\arcmin\,$-$\,4.7\arcmin or 15.0\,$-$\,20.5\,kpc). They are well defined and large to the north, thinner to the west and more diffuse to the south (see Fig.\,\ref{optvst}). The east region shows very diffuse emission and no clear sign of the pseudo-ring structure, making the latter a partially closed pseudo-ring. Despite the SB limit reached, the numerous stars in the field and their PSFs may play a relevant role by overlapping with any fainter emission at such low SB, mimicking non-existent extragalactic stellar traces \citep{trujillo16}. In particular, this occurs in the eastern region where a few bright stars are located.
However, the even deeper SB limit reached with the VST image has two immediate implications: one, the definition of certain regions of the pseudo-ring are greatly improved and two, the Galactic cirrus starts to become clearly visible at 28.5\,mag\,arcsec$^{-2}$, hindering the detection of features beyond the pseudo-ring at SBs fainter than this level (see Sect.\,\ref{sec:cirrus}).


\subsection{Disc and pseudo-ring relative orientation}
\label{sec:isophotes}

A visual inspection of the CAHA2.2m optical image suggested an apparent misalignment between the pseudo-ring and the galactic disc. In order to quantify it, we performed elliptical fittings to the pseudo-ring structure as well as to the isophotes of the galaxy from 20.2 to 26.4\,mag\,arcsec$^{-2}$ after removing the signatures of the close bright stars to avoid biased fittings. 

The fittings of the innermost regions of the galaxy ($\mu_{Cous\ R}$\,=\,24.0\,mag\,arcsec$^{-2}$ or brighter) were not reliable because of the strong influence of the spiral arms. Moreover, bright close stars contaminate the outer regions (fainter than $\mu_{Cous\ R}$\,=\,24.0\,mag\,arcsec$^{-2}$). Even after removing them, too few points are left making reliable fittings difficult.

However, the optical images clearly showed the centre of the galaxy (error below 1\arcsec). After fitting the pseudo-ring we found a shift of 12$\arcsec$ ($\sim$1.2\,kpc, the approximate length of the bar) between the centres of the pseudo-ring fitting and the disc and its orientation was PA$_{pseudo-ring\ fit}$\,=\,21.5\degree, similar to the PA of the major axis of the galaxy (PA$_{maj}$\,=\,20\degree). We also de-projected the image assuming a disc inclination of $i_{CIG\,96}$\,=\,46.59$\deg$ to confirm whether the pseudo-ring may be oval or in a different plane from the disc. We found the flattening or ellipticity of the pseudo-ring is of 0.04\,$-$\,0.05\%, that is, practically circular, suggesting it to be slightly oval if seen at almost the same inclination as the inner disc of the galaxy.

%
\begin{figure}
\centering
\includegraphics[width=\hsize]{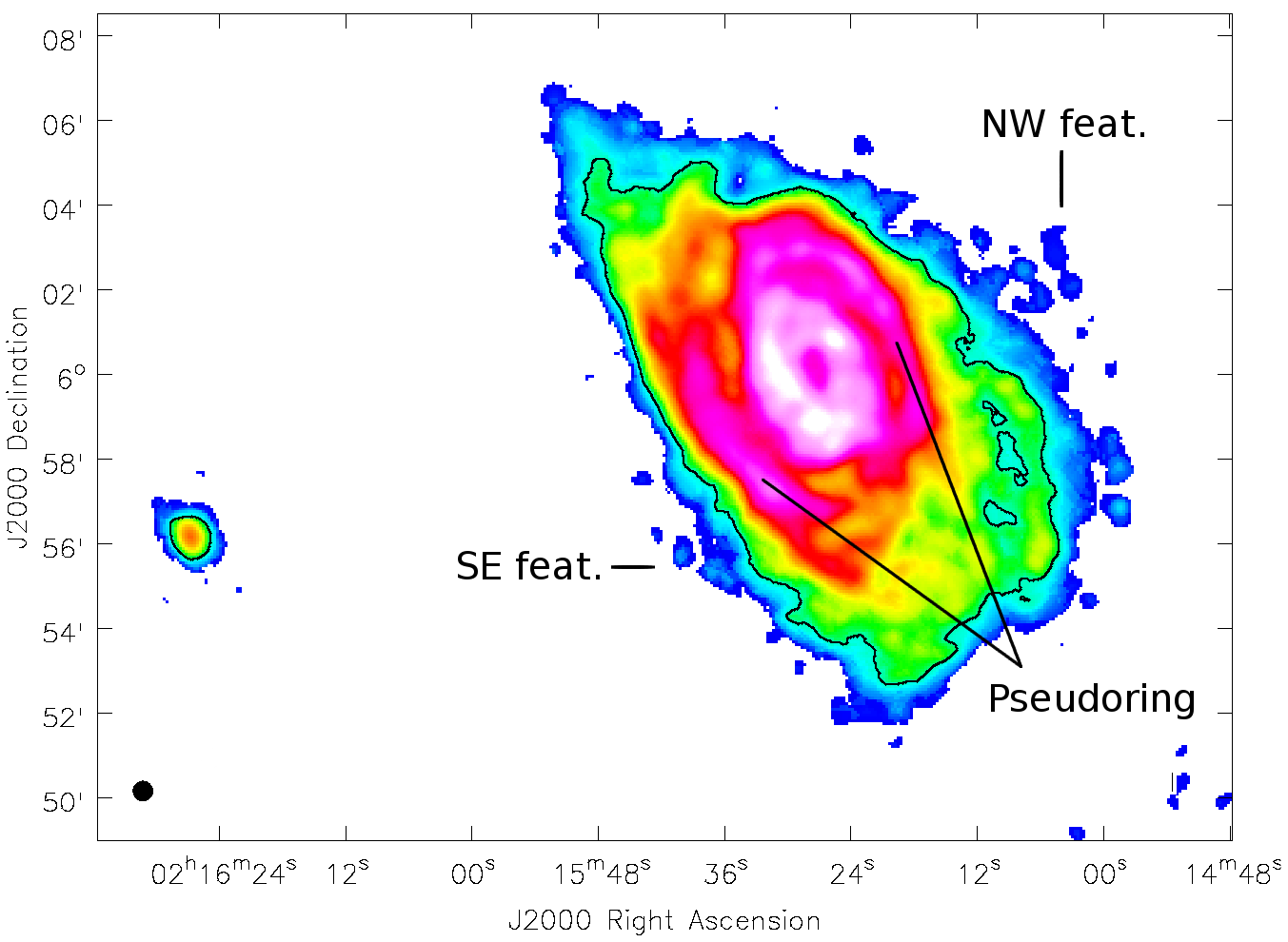}
   \caption{HI integrated intensity map of CIG\,96 and its companion after a 3.5$\sigma$ blanking (see Sect.\,\ref{sec:lownoise}). We identify the NW and SE HI features mentioned in Sect.\,\ref{sec:chmap} as well as the HI emission of the pseudo-ring. The black contour represents the column density of $N_{HI} = 8.7\times10^{20}$ cm$^{-2}$ (5$\sigma$) reached by \cite{espada11a}. The black circle at the bottom left represents the beam size of 28\arcsec $\times$ 28\arcsec.}
      \label{mom0}
\end{figure}

%

%
\begin{figure}
\centering
\includegraphics[width=\hsize]{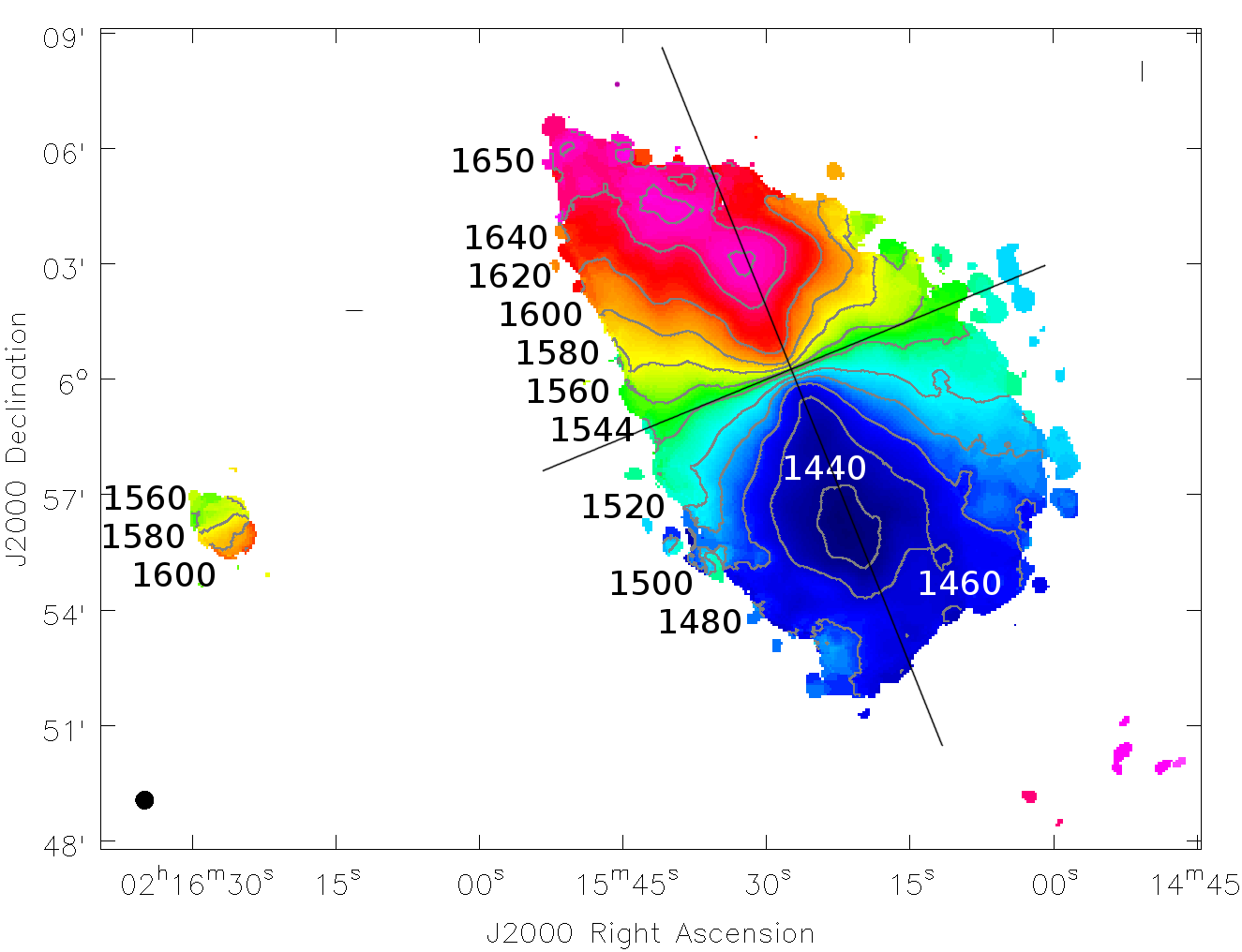}
   \caption{HI velocity field map of CIG\,96 and its companion after a 3.5$\sigma$ blanking (see Sect.\,\ref{sec:lownoise}). The black lines indicate the orientation of the major and minor axis (PA$_{maj}$ = 20$\degree$ and PA$_{min}$ = 110$\degree$, respectively) along which the position-velocity cuts have been performed (see Fig.\,\ref{pvcuts}). Grey contours represent the indicated velocities in km\,s$^{-1}$. The black circle at the bottom left represents the beam size of 28\arcsec $\times$ 28\arcsec.}
      \label{mom1}
\end{figure}

%
\begin{figure}
\centering
\includegraphics[width=\hsize]{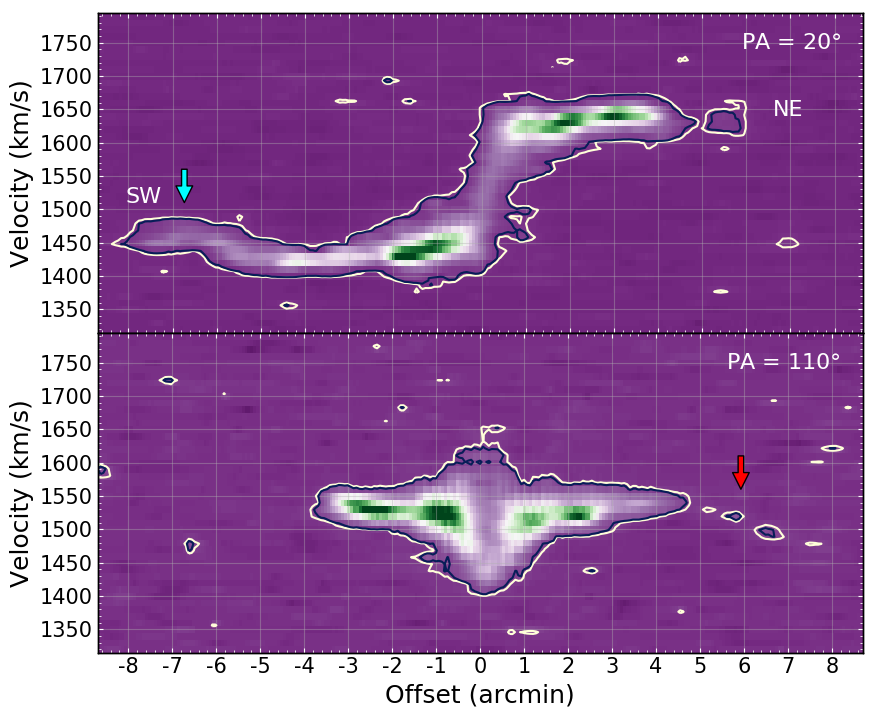}
   \caption{Position-velocity cuts along the major axis (\textit{top panel}, PA\,=\,20$\degree$) and minor axis (\textit{bottom panel}, PA\,=\,110$\degree$) of CIG\,96 HI cube. The column density is $N_{HI}\ (1\sigma) = 0.24\times10^{20}$ cm$^{-2}$ and the white and black contours correspond to 3.5$\sigma$ and 5$\sigma$, respectively. The cyan arrow points to the SW region where the velocity increases by approximately 30 $-$ 40 km\,s$^{-1}$ (see Sect.\,\ref{sec:HImoments}). The red arrow points to the NW HI feature, the clumpy structure detected visible in channels 16 to 23 of the HI cube (see Sect.\,\ref{sec:chmap}). As a reminder, the beam resolution is of 28\arcsec $\times$ 28\arcsec.}
      \label{pvcuts}
\end{figure}



\subsection{CIG\,96 colour index image and optical features}
\label{sec:pseudocolor}

We analyse here the colour index image of CIG\,96 and the distribution along the pseudo-ring (further discussed in Sect.\,\ref{sec:feats96}) via CAHA1.23m B and R images converted to SDSS\,$g$ and $r$ magnitudes, respectively (see Sect.\,\ref{sec:123data}).



\begin{figure*}
\resizebox{\hsize}{!}
 {\includegraphics[scale=0.5]{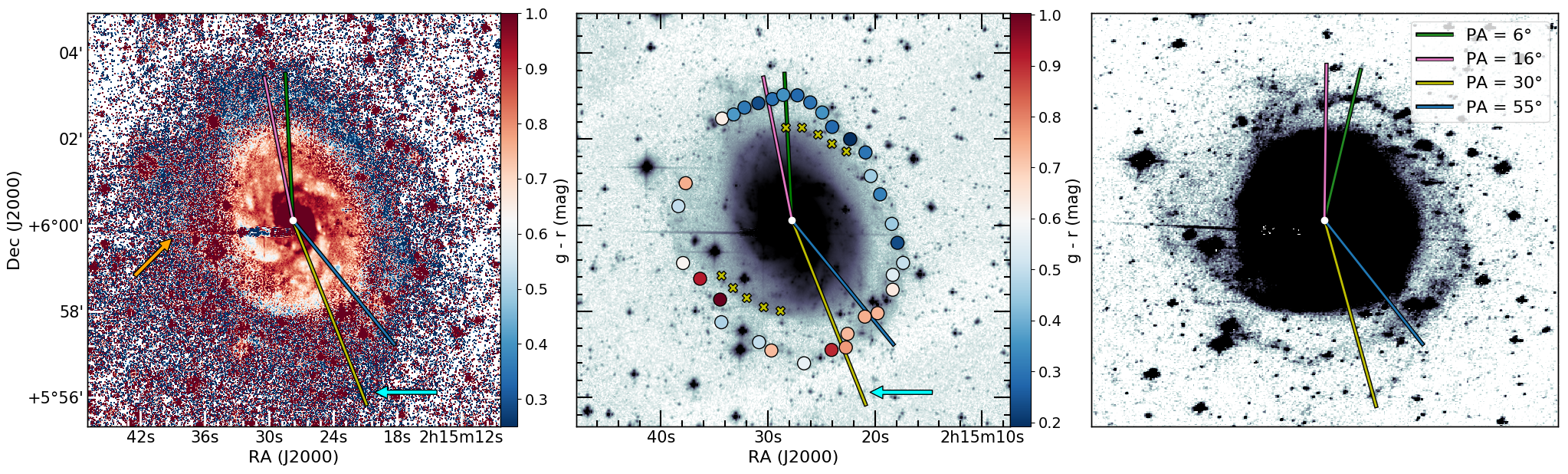}}
\caption{\textit{Left panel}: SDSS $g-r$ colour index image. 
Orange arrow points to the eastern arc of the pseudo-ring. Cyan arrow points to an optical feature to the south. Blue and red tones indicate colour index in magnitudes. The pink, blue and green lines indicate the directions (or PA) used to compute the four radial profiles discussed in Sect.\,\ref{sec:radcuts}; they are repeated in the other panels. \textit{Central panel}: 33 circular apertures of 1.25 kpc radius (12.7\arcsec) located over star-free areas of the pseudo-ring, represented over the CAHA2.2m and CAHA1.23m combined optical image. A bluer/redder region indicates a bluer/redder $g-r$ colour index. The cyan arrow points again to the southern feature. Yellow marks indicate the connecting regions between the pseudo-ring and the inner parts of the galaxy. \textit{Right panel}: De-projected SDSS\,$g$ CAHA1.23m image of CIG\,96; the lines of the radial profiles have been reoriented to preserve the correct PA.}
\label{grpseudocuts}
\end{figure*}


As a reference for the colour index values plotted in Fig.\,\ref{grpseudocuts} left and central panels, we indicate the boundaries of the blue and red clouds from the SDSS $g-r$ optical colour$-$magnitude diagram. In particular we show the Green Valley interval of $\left(g - r\right)_{G.V.} = 0.60 - 0.75$ mag as defined by \cite{walker13} following the colour analysis by \cite{strateva01}.

Qualitatively, we also note three striking features from the $g-r$ and optical images (see Fig.\,\ref{grpseudocuts}, left and central panels). The first feature is a diffuse arc in the east side of the pseudo-ring that almost closes it from north to south (Fig.\,\ref{grpseudocuts}, left panel, orange arrow); it is barely detectable (below $\sim$ $1.2\sigma$) in any individual image further than a diffuse emission due to the heavy contamination of nearby stars. The second structure is also barely detectable (below $1.2\sigma$) in any individual image despite there being no significant contamination by close stars in this region. It is located beyond the southern region of pseudo-ring, approximately 30\,kpc ($\sim$5\arcmin) from the galaxy centre (Fig.\,\ref{grpseudocuts}, left and centre panels, cyan arrow). The third structure is indicated with yellow crosses in the central panel of Fig.\,\ref{grpseudocuts}. This double structure has a SB of $\sim$26.0\,mag\,arcsec$^{-2}$seems to connect the northern and southern inner parts of the galaxy with the western and eastern sides of the pseudo-ring, respectively.

Both in our VST image and in the DECaLS DR5 image, we detect a faint elongated (approximately $\sim$1\arcmin\ long) and diffuse structure to the NE of CIG\,96 (coordinates RA\,=\,$2^{h}15^{m}58.286^{s}$, DEC\,=\,+6\degree 04\arcmin 39.15\arcsec). It is located close to a bright star and barely a few kiloparsecs beyond the field of view covered by our $g-r$ image. As described further in Sect.\,\ref{sec:cirrus}, this structure lies on a region with a noticeable amount of background emission, mostly due to Galactic cirrus, and could therefore be part of it. However, we cannot rule out that this feature might be a tidal stream or a tidal disruption dwarf.


\subsubsection{Pseudo-ring colour index distribution}
\label{sec:colordistribution}

\begin{figure*}
\resizebox{\hsize}{!}
 {\includegraphics[scale=0.5]{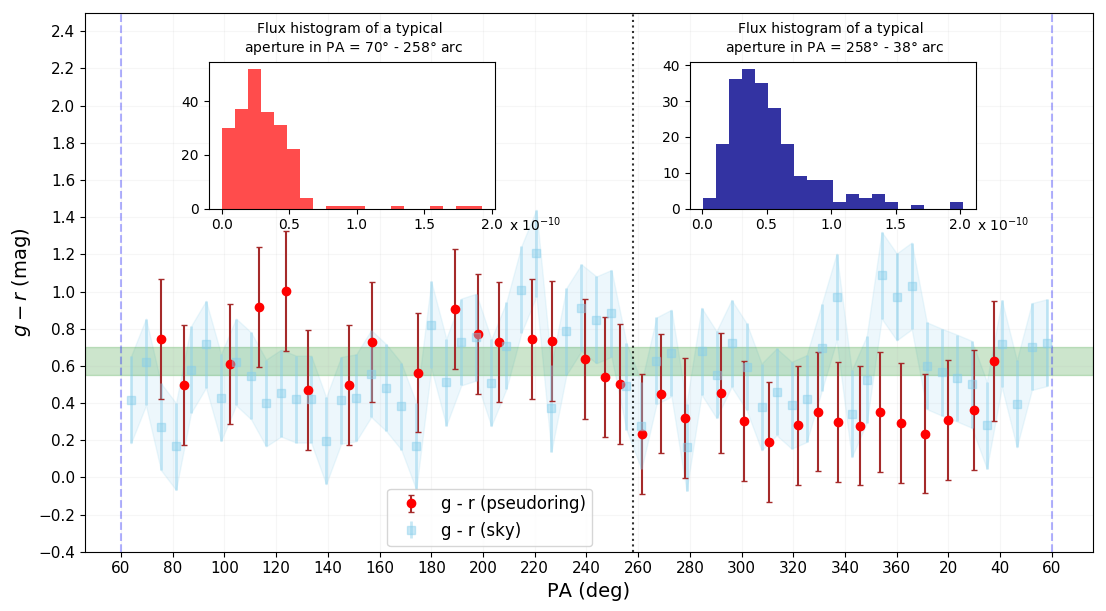}}
\caption{SDSS $g-r$ colour index vs. PA along the pseudo-ring. The red dots represent the $g-r$ values obtained from the de-projected images. They are obtained by dividing $g$ median flux to the corresponding $r$ median flux of each aperture and converting these results to magnitudes. The pale blue dots correspond to the $g-r$ colour index measured at a distance of  $r$ = 29.5 kpc ($\sim$5\arcmin) on the sky. The green stripe sets the Green Valley interval that separates the red cloud ($g-r>0.75$ mag) from the blue cloud ($g-r<0.60$ mag) as defined in \ref{sec:pseudocolor}. The embedded figures correspond to typical flux $\times$10$^{-10}$ histograms for two apertures from the SE region (PA\,=\,70$\degree$ $-$ 258$\degree$ and the NW region (PA\,=\,258$\degree$ $-$ 38$\degree$) separated by the vertical doted grey line.}
\label{pseudoapers}
\end{figure*}

We have studied the azimuthal variation of the colour index along the pseudo-ring by determining its median value in 33 circular non-overlapping apertures distributed in foreground star-free regions along its extent as shown in the central panel of Fig.\,\ref{grpseudocuts}, except for the NE region (PA in the range 38$\degree -$ 70$\degree$) due to the lack of reliable optical emission in this arc. We defined the apertures over a de-projected image of CIG\,96. For a better visualization, we have kept their spatial location and circular shape in the image presented in the previous figure, which is not de-projected.
In order to discard any colour index changes in the pseudo-ring due to a gradient in the sky level, we determined the sky colour index of 62 regions set farther than the pseudo-ring, covering 360\degree around CIG\,96 and free of bright stars. These apertures show $g-r$ values between $\sim$0.2 and $\sim$1.2. In Fig.\,\ref{pseudoapers} we show the $g-r$ colour index distribution of all regions according to their PA and we find no colour index correlation between the apertures from the pseudo-ring and those from sky.
However, as anticipated in the central panel of Fig.\,\ref{grpseudocuts}, we find a colour index change in between two PA ranges of the pseudo-ring. The 17 apertures of the SE arc (within PA\,=\,70$\degree -$258$\degree$) show a median colour of $g-r = 0.73$ mag ($st.dev. = 0.15$ mag), that is, a redder colour. Contrarily, the remaining 16 apertures of the NW arc (within PA\,=\,258$\degree -$38$\degree$) show a median value of $g-r = 0.31$ mag ($st.dev. = 0.11$ mag), that is, a bluer colour, making the difference between the two regions of $g-r \simeq$ 0.4 mag.



\subsubsection{Radial cuts}
\label{sec:radcuts}

In order to compare the colour of the disc with the immediate pseudo-ring regions we computed radial profiles from individual $g$ and $r$ images. The right panel of Fig.\,\ref{grpseudocuts} shows the de-projected $g$ image of CIG\,96 together with the lines along which those were calculated. 

\begin{figure}
\centering
\includegraphics[width=\hsize]{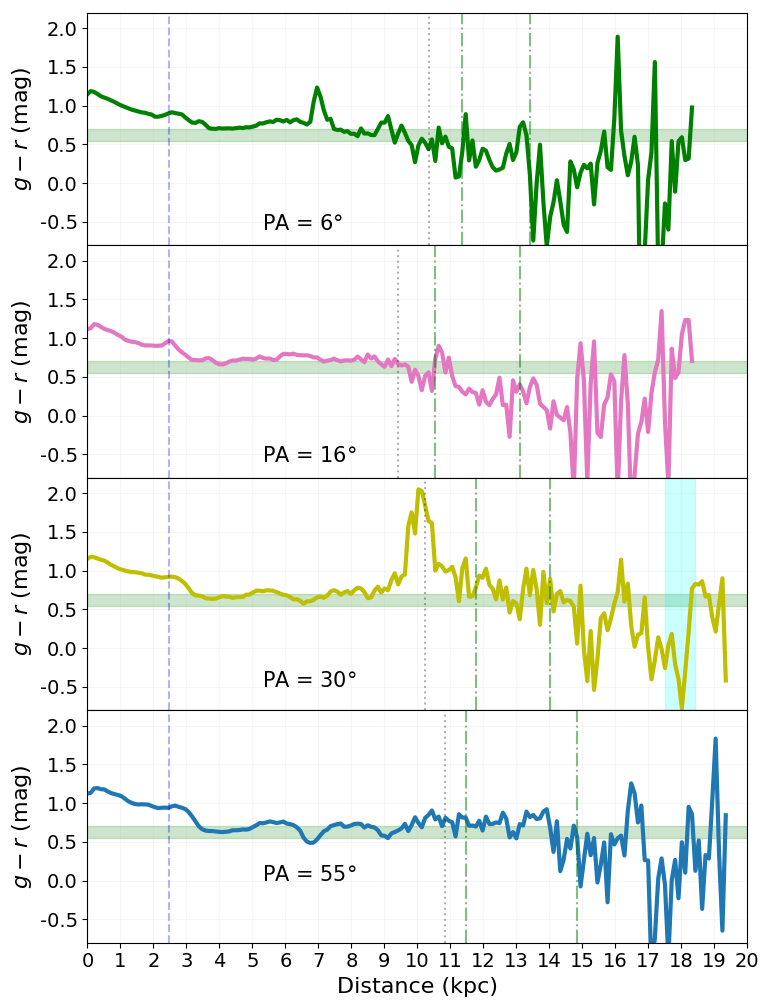}
   \caption{Radial profiles obtained along four different orientations with PA of 6$\degree$, 16$\degree$, 30$\degree$ and 55$\degree$ (shown in Fig.\,\ref{grpseudocuts}, right panel) at the top, top-centre, bottom-centre and bottom panels, respectively. The horizontal green stripe represents the Green Valley in SDSS $g-r$ (see Sect.\,\ref{sec:pseudocolor}. The bulge, disc and pseudo-ring limits are measured at 26.8\,mag\,arcsec$^{-2}$. The bulge limit (2.5 kpc) and the disc limit are shown as the light blue dashed and black dotted lines, respectively. The pseudo-ring variable inner and outer limits are defined in each panel by the green dot-dashed lines. The light cyan band in the PA\,=\,30$\degree$ panel sets the location of the optical feature to the S marked as a cyan arrow in the left panel of Fig.\,\ref{grpseudocuts} and a magenta ellipse in Fig.\,\ref{grHI}.}
\label{radprof}
\end{figure}

These profiles are shown in Fig.\,\ref{radprof}, where the bulge (the first 2.5 kpc, \citealt{espada11a}), disc and pseudo-ring radii are marked as well. We selected the orientations due to the different structures crossed: disc, dust regions, arms, star-forming regions and thicker/thinner regions of the pseudo-ring. The profiles were then computed at PA of 6$\degree$, 16$\degree$, 30$\degree$ and 55$\degree$ and we will refer to them as PA6, PA16, PA30 and PA55, respectively. 

To present the main results that these profiles yield, we have used a SB of 26.8\,mag\,arcsec$^{-2}$ in the SDSS\,$r$ band. At this depth, the disc size varies in a range of $R_{disc}$\,=\,9.5$-$\,11\,kpc, depending on the PA.

The gap between the disc and pseudo-ring is not constant either: in the regions where the pseudo-ring and the disc are well resolved, the gap has an approximate width of $\simeq$1\,kpc. However, in regions where both the disc and pseudo-ring have a more diffuse emission, they prevent any reliable estimation of this separation.

The gap width, as well as its uncertainty, has a connection to the pseudo-ring dimensions: the more defined regions of the pseudo-ring have a width of $w_{pseudo-ring}\simeq$\,2\,kpc but it may rise up to $\sim$4\,kpc in some diffuse regions being hardly distinguishable from the disc.

Profile PA6 (green) shows red colours along the disc relative to the limit defined by the Green Valley strip. The peak at $\sim$7\,kpc corresponds to a foreground star ($m_{r\ SDSS}$\,=\,19.65\,mag). The pseudo-ring shows blue colours in most of its extent along this PA, matching the star-forming region ($\sim$0.7\,kpc size) present in this section of the cut, centred at a radius of approximately 12.5\,kpc. The colour difference of the disc and the pseudo-ring at this PA is $\sim$0.4\,mag.


Profile PA16 (pink) also shows the difference in colours between the disc and the pseudo-ring. The disc shows a stable red colour throughout its whole extension ($g-r \simeq$\,0.7\,mag). However, the pseudo-ring shows a colour gradient from $g-r \sim$0.7 to 0.1\,mag approximately, hence most of the pseudo-ring has blue colours. This profile was also aimed towards a large star forming region of $\sim$1.5\,kpc radius in the pseudo-ring and located at an approximate distance of 12\,kpc, so such a blue colour is expected. However, there is no apparent cause for the colour change.

Profile PA30 (yellow) shows a uniform disc colour within or right on the red edge of the Green Valley ($g-r \simeq$\,0.75\,mag) consistent with the rest of the profiles. There is an exceptionally red peak at 10.5\,kpc that, unlike in the case of PA6 (produced by a star), is the result of a region with large quantities of dust. The orientation of the previous profiles missed these dusty inner regions of the galaxy, easily visible in the left panel of Fig.\,\ref{grpseudocuts}, left. PA30 crosses the pseudo-ring through an area of diffuse emission and the redder colour is consistent throughout its extension. The orientation of this profile was chosen to obtain also the colour of the southern feature of $\sim$1\,kpc in width located at $\sim$18\,kpc indicated with a red vertical stripe (also marked with a cyan arrow in Fig.\,\ref{grpseudocuts}, left panel). Despite the fact that the feature is surrounded by the sky, its location and the surrounding 0.5\,kpc show a clearly blue colour. We have not considered this feature as part of the pseudo-ring so its width estimation remains between 1.5 and 3.5\,kpc and its radius $\sim$14\,kpc.

Profile PA55 (blue) shows a different behaviour along the disc. The mean colours are bluer than along the previous profiles; two regions that correspond to where the arms are crossed show very blue colours. 
This profile was selected to observe a much more diffuse and broad region of the pseudo-ring (width up to $\sim$3.5\,kpc). As in the disc, the pseudo-ring colour along this orientation is not homogeneous but it shows red colours ($g-r \sim$\,0.8\,mag) throughout most of its width. The farthest part of the pseudo-ring shows a steep change towards bluer colours, making it difficult to decipher whether it is an artifact of the sky or an existing structure with similar colour.



\subsection{Colour index and HI column density in the pseudo-ring}
\label{sec:colorandHI}

\begin{figure}
\centering
\includegraphics[width=\hsize]{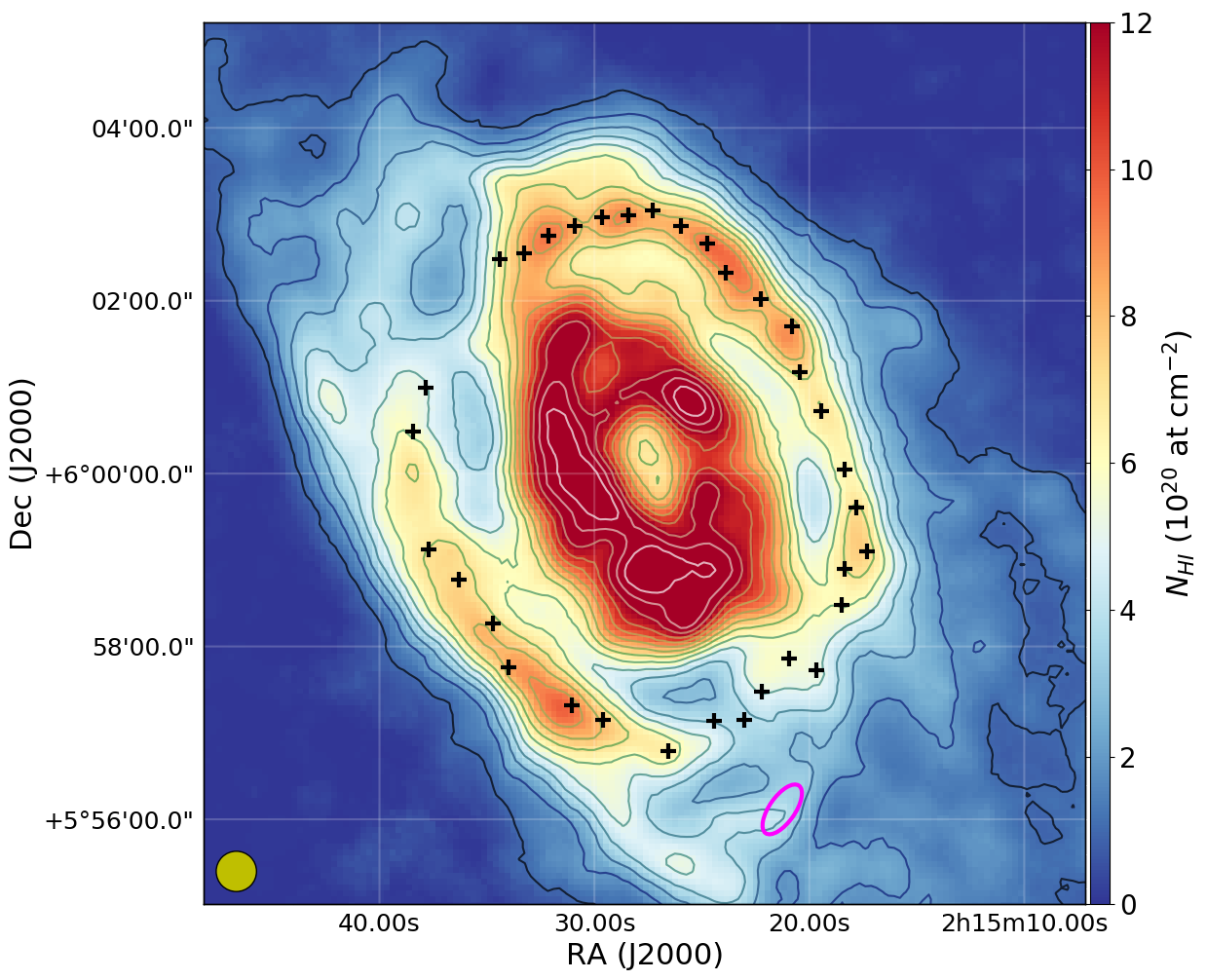}
\caption{Central 53$\times$53 kpc (9$\times$9 arcmin) of the integrated HI emission map of CIG\,96. Column density is indicated with a colour gradient. Contours indicate 1, 2, 3, 4, 5, 6, 7, 8, 9, 10, 11, 12, 13 and 14 $\times$ 10$^{20}\ at\ cm^{-2}$. The black crosses indicate the central position of the 33 apertures used to measure the colours of the pseudo-ring (see Sect.\,\ref{sec:colordistribution}). The magenta ellipse indicates the position of the southern feature indicated with a cyan arrow in Fig.\,\ref{grpseudocuts}, left panel, the $g-r$ colour index image. The yellow circle at the bottom left indicates the HI image synthesized beam of 28$\arcsec\times28\arcsec$.}
\label{HIpseudomarks}
\end{figure}

The black crosses of Fig.\,\ref{HIpseudomarks} show the location of the apertures of the pseudo-ring on top of the HI 0$^{th}$ moment map. We find a remarkable spatial correlation between the optical pseudo-ring and the HI distribution, in agreement with \cite{espada11a}. In Fig.\,\ref{HIpseudomarks} we indicate with a magenta ellipse the spatial location of the optical southern feature (shown in Fig.\,\ref{grpseudocuts} left panel with a cyan arrow). It is too distant from the pseudo-ring ($\sim$4.1 kpc) as to confirm that both have a physical link and, unlike other star-forming regions of the pseudo-ring, we find no increase of the $N_{HI}$ in this region.

\begin{figure}
\centering
\includegraphics[width=\hsize]{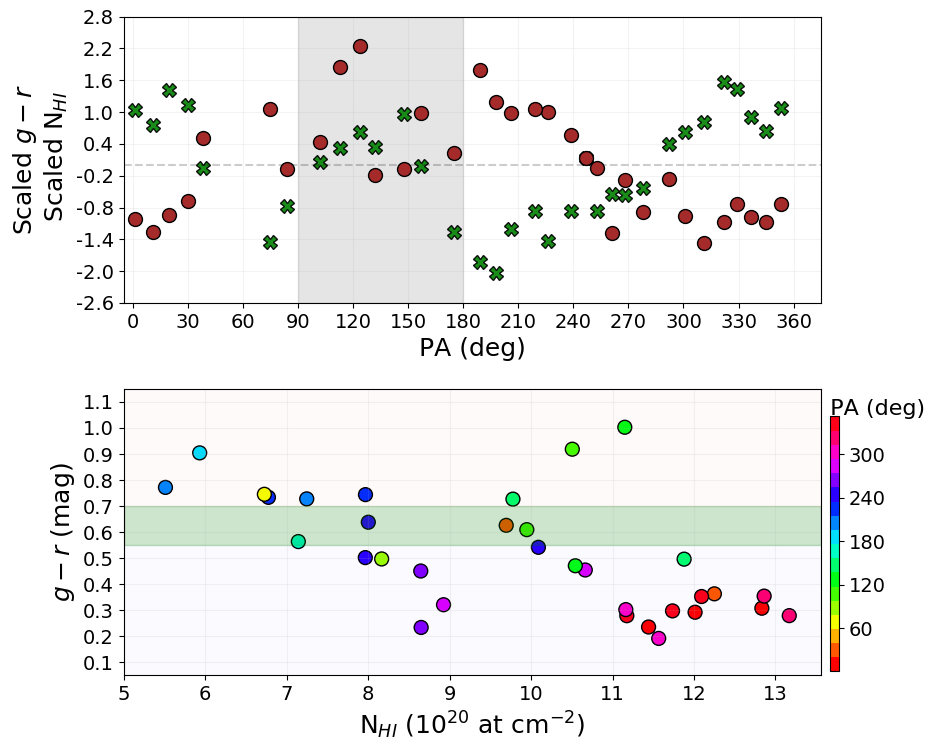}
\caption{\textit{Top panel}: $g-r$ (brown circles) and $N_{HI}$ (green crosses) scaled (mean subtracted, sigma divided) values in each of the 33 apertures. A grey dashed line has been drawn at $g-r$ = 0 for reference. Redder colours and higher $N_{HI}$ are positive in this figure. \textit{Bottom panel}: $g-r$ median colour index vs. $N_{HI}$ measured in the 33 apertures traced over the pseudo-ring. The green horizontal stripe represents the Green Valley in SDSS $g-r$.}
\label{grHI}
\end{figure}

We have performed a detailed comparison between the pseudo-ring colour index $g-r$ and $N_{HI}$ for each selected aperture. With this aim, we have scaled each one of them by subtracting the mean value of the 33 apertures and dividing them by their sigma value (Fig.\,\ref{grHI}, top panel). We observe an anticorrelation between $g-r$ and $N_{HI}$ scaled values within PA\,=\,180$\degree -$ 40$\degree$, i.e. bluer colours correspond to larger column densities. It is only broken in the range PA\,=\,90$\degree -$ 180$\degree$, approximately, probably due to less reliable $g-r$ measurements in this side of the pseudo-ring, the most diffuse region. The anticorrelation is also confirmed in the bottom panel of Fig.\,\ref{grHI} where we show $g-r$ as a function of $N_{HI}$: most of the bluer areas, mainly located in the NW side of the pseudo-ring (PA\,=\,260$\degree -$40$\degree$) show column densities of 8.5\,$-$\,13.5\,$\times$ 10$^{20}$ cm$^{-2}$, higher than most of the redder ones which show much lower levels instead. These results are discussed further in Sect.\,\ref{sec:feats96}.



%


\subsection{Optical characteristics of the companion}
\label{sec:companion}

\begin{figure}
\centering
\includegraphics[width=\hsize]{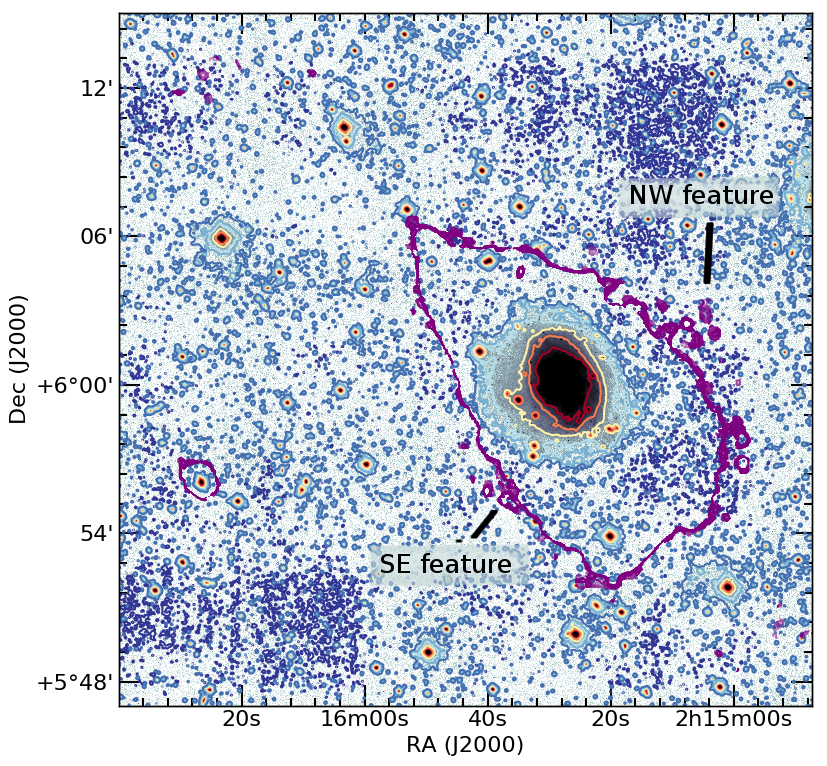}
   \caption{VST image in SDSS\,$r$ band of a 28$\arcmin\times28\arcmin$ region covering both CIG\,96 and its companion NGC\,864\,COM1 (indicated by the magenta contour to the east of the image). Optical contours are set in 24.0, 25.0, 26.0, 27.0, 27.5, 28.0 and 28.4\,mag\,arcsec$^{-2}$ (SDSS\,$r$), smoothed with a Gaussian kernel of 11 pix radius. The purple contour marks the HI column density limit of the zeroth moment at 0.6$\times$10$^{20}\,cm^{-2}$, as in Fig.\,\ref{mom0opt}.}
      \label{cig96companion}
\end{figure}

We aimed to observe any possible optical structures connecting CIG\,96 and its companion (see Sect.\,\ref{sec:cigcomp}). Given that the CAHA images have a field of view of 12$\arcmin \times$12$\arcmin$ (approximately 71\,kpc$\times$71\,kpc), it is only possible to studying any potential optical connection between the two galaxies with the VST 1$\degree\times$1$\degree$ image, as it provides continuous coverage across the $\sim$90\,kpc separation between them. Figure\,\ref{cig96companion} shows a 10\arcmin$\times$10\arcmin\ image centred between the two galaxies. At the current optical SB limit and in agreement with the HI map, we detect no sign of any stellar feature tracing any direct interaction between CIG\,96 and its companion.

Focusing on its companion, our VST image shows that it consists of an elongated structure oriented with a PA of 35\degree plus a spheroidal component. Although the HI resolution of our data prevents us from separating both optical components, the HI emission is slightly elongated along the same PA, within the resolution of our HI data (28\arcsec\,$\times$\,28\arcsec, see Fig.\,\ref{mom0}). Moreover, the HI kinematics shows the same orientation and is quite regular (within the limits given by the beam smearing), especially in the velocity range of 1560 to 1610\,km\,s$^{-1}$, where the main HI emission of the galaxy is found (see Fig.\,\ref{mom1}).

In order to determine whether the optical image of NGC\,864\,COM01 is compatible with a disc$-$bulge system, we proceeded in two steps. We inspected the model and residuals provided by DECam Legacy Survey (DECaLS DR5, \citealt{schlegel15}) and the model shows a good fit to the spheroidal component of the galaxy, with the residuals suggesting a blue and close to edge-on disc with a similar PA to the HI velocity field, as indicated above. Hence, from a morphological point of view it is compatible with a spiral galaxy with an Sa - Sb type.

\begin{figure}
\centering
\includegraphics[width=\hsize]{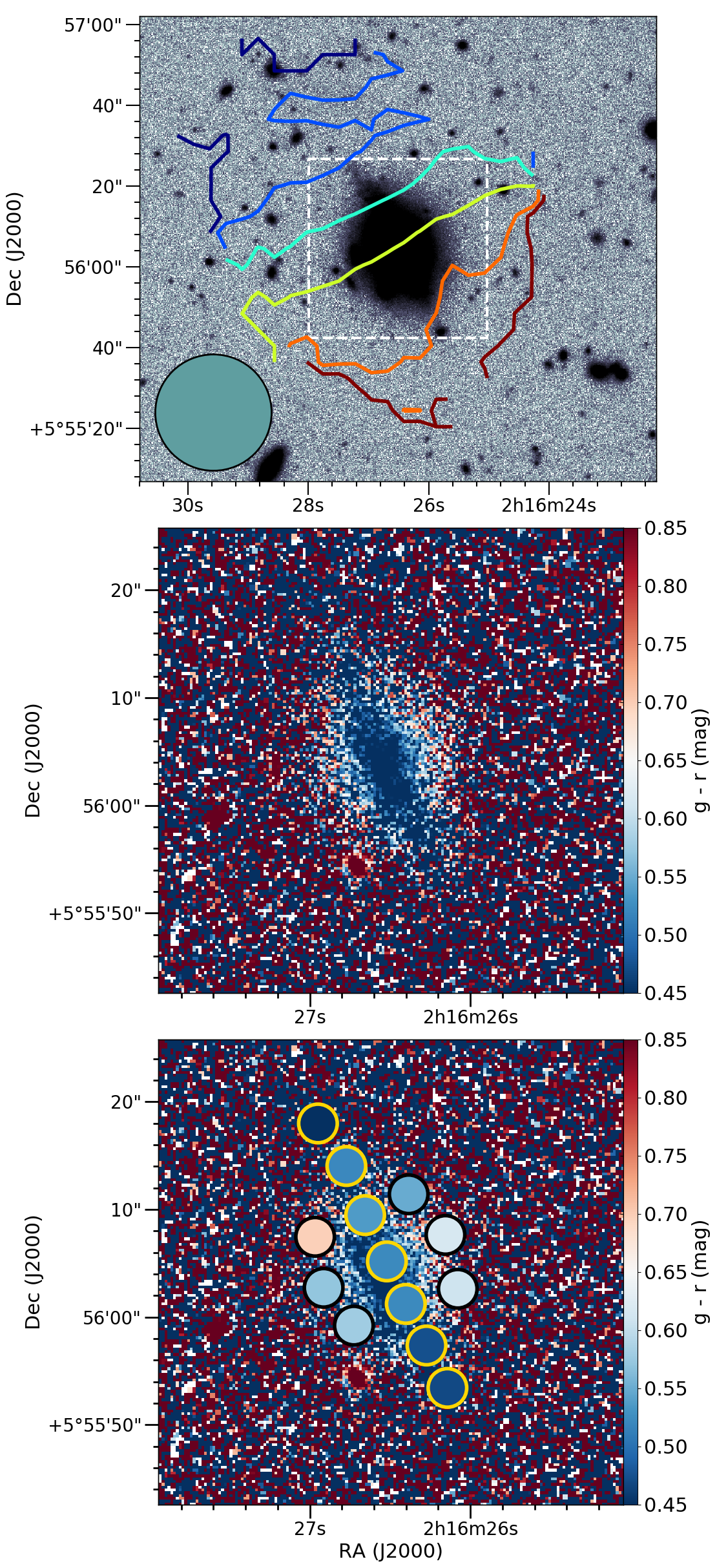}
\caption{\textit{Top panel}: Velocity contours of the blanked HI datacube from 1560 to 1610\,km\,s$^{-1}$ (in steps of 10\,km\,s$^{-1}$) on top of the VST image of the companion of CIG\,96. The beam size of the HI data is indicated as a 28\arcsec$\times$28\arcsec grey disc in the bottom left corner of the image. The white frame represents the central 45\arcsec$\times$45\arcsec of the SDSS\,$g-r$ image of the bottom panel.
\textit{Central panel}: SDSS\,$g-r$ colour index image of the companion of CIG\,96 built from the corresponding SDSS\,$g$ and $r$ images from DECaLS\,DR5 survey.
\textit{Bottom panel}: Same image as in the central panel with the apertures used to measure the $g-r$ colour index. All apertures have a 1.75\arcsec radius. The yellow-edge apertures trace the colours of the elongated feature while the black-edge apertures show the colours of the galaxy.}
\label{companionHIgr}
\end{figure}

As a next step we decided to perform a similar study to the analysis of the pseudo-ring of CIG\,96, aiming to determine the colour of each of the two components and whether they are consistent with values found for spheroidal components and discs in spiral galaxies. In particular, we used SDSS\,$g$ and $r$ images from the DECaLS survey whose exquisite seeing allowed to produce a $g-r$ colour index image and measure the colours in the sky (median value is 0.49$\pm$0.07\,mag) as well as in certain apertures, as shown in Fig.\,\ref{companionHIgr}, central and bottom panels. The measures in these apertures show that the elongated structure is bluer (0.33$-$0.45\,mag) than the spheroidal one, which is redder (0.55$-$0.70\,mag, right over the Green Valley range, see Fig.\,\ref{pseudoapers}).

Lastly, we checked whether Sa-Sb galaxies exist with masses similar to that of the CIG 96 companion. We have used the value for the dynamical mass of NGC\,864\,COM01 as determined in Sect.\,\ref{sec:sblimit} ($\sim$10$^{9}$\,M$_{\odot}$ which, as explained there, is a lower limit to the real one). According to \cite{nair10} there exist Sa and Sb galaxies within such a mass range. Hence all these characteristics lead us to favour the possibility of NGC 864 COM01 being an Sa or Sb galaxy.

\begin{figure*}
\resizebox{\hsize}{!}
 {\includegraphics[scale=0.5]{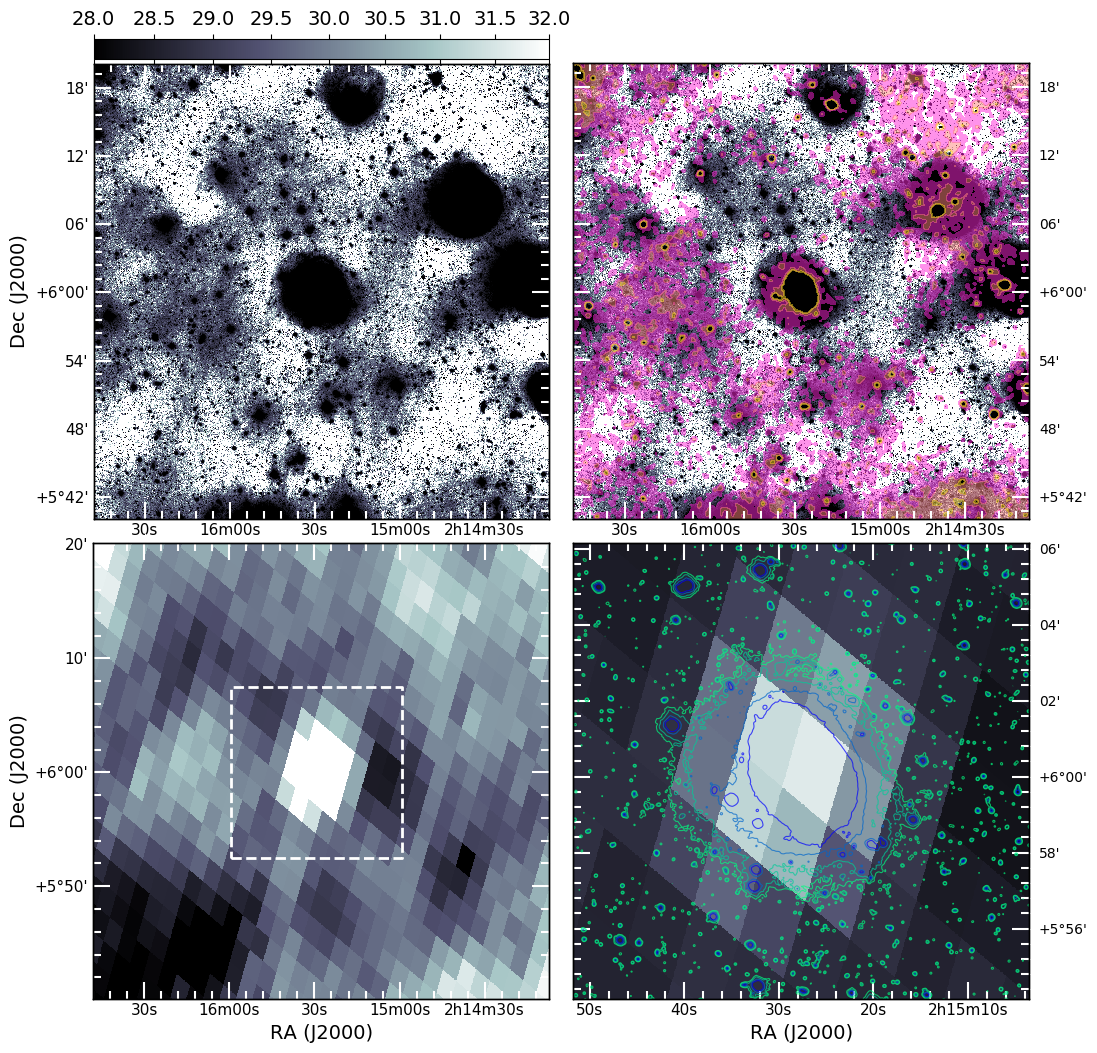}}
   \caption{\textit{Top left panel}: Central $40\arcmin\times40\arcmin$ of the VST optical image in SDSS\,$r$ band of CIG\,96. The SB is indicated on top of the figure in\,mag\,arcsec$^{-2}$ (SDSS\,$r$ band) favouring the levels that enhance the Galactic cirrus located in the field of our galaxy, which show up from $\mu_{r\ SDSS} \simeq$ 28.5\,mag\,arcsec$^{-2}$. \textit{Top right panel}: Background, the same $40\arcmin\times40\arcmin$ VST optical image as in top left panel; foreground, overlay of WISE band 3 (12 $\mu m$) smoothed image contours (Gaussian kernel of 21 pixel radius). \textit{Bottom left panel}: $40\arcmin\times40\arcmin$ Planck 857 GHz image of CIG\,96. The dashed white square indicates the approximate area selected for the bottom images. \textit{Bottom right panel}:  Background, $12\arcmin\times12\arcmin$ Planck 857 GHz image; foreground, 25.0, 26.0, 27.0, 27.5\,mag\,arcsec$^{-2}$ smoothed (Gaussian kernel of 7 pixel radius) optical contours from VST image (SDSS\,$r$ band).}
      \label{cirrus}
\end{figure*}


\subsection{Cirrus}
\label{sec:cirrus}

The possible large PSFs from close stars (whose radii reach several arcminutes, \citealt{trujillo16}) and the necessity of a precise sky subtraction are key to discerning what may be artifacts from actual faint stellar components. However, there is another limiting factor at the current optical SB: the emission due to the presence of Galactic cold dust (\citealt{sandage76}, confirmed with IRAS data by \citealt{neug84}) known as cirrus in the Milky Way. This cirrus may cover large areas in the sky and, depending on the region, may start to be especially relevant from $\mu_{r\ SDSS} \simeq$ 25.5 $-$ 26.0\,mag\,arcsec$^{-2}$ in the optical regime \citep{duc15}. 

As described in Sect.\,\ref{sec:planckwise}, we have used Planck857 and WISE3 images to identify the presence of cirrus in the field of CIG\,96. Their optical resolutions are 4.2$\arcmin$ and 6.5$\arcsec$, respectively, that is, approximately 242 and 6 times lower than the 1.04 arcsec pix$^{-1}$ resolution of our CAHA images or 1260 and 30 times lower than our VST image, respectively.

The Planck857 image is a good indicator of cirrus; however, we cannot extract reliable conclusions in the field of CIG\,96 due to its low spatial resolution. We used the central $40\arcmin\times40\arcmin$ of the VST image to inspect the cirrus (see Fig.\,\ref{cirrus}, top left panel). It shows that this area is populated with scattered emission visible from $\sim$28.5\,mag\,arcsec$^{-2}$. While this image is not enough to conclude whether that emission is associated with cirrus or not, we can confirm it does not show up in any other observation considered in this work.

In the Planck857 image (Fig.\,\ref{cirrus}, bottom left panel), a $\sim$2$\sigma$ peak of emission (where $\sigma$ is the $rms$ measured on the clean SE region) is visible in the central pixels where the galaxy is located. In the surrounding area and close to the noise level, there are extended areas to the east, northwest and southwest of CIG\,96 which seem to match some of the emission observed with the VST image at 28.5\,mag\,arcsec$^{-2}$. 

The lack of cirrus structures brighter than 28.5\,mag\,arcsec$^{-2}$ suggests they provide scarce (if any) contamination at brighter levels in our VST image, hence we set our detecion limit at 28.4\,mag\,arcsec$^{-2}$ in SDSS\,$r$ band. From the opposite point of view, the low resolution of Planck857 makes it pointless to use such an image to find any cirrus structure in our optical image (Fig.\,\ref{cirrus}, bottom right panel). Should these exist, a positive matching between our VST and Planck 857 images would require extremely large and bright structures, easily detectable in both images; however, we do not find such large structures, preventing the use of the Planck857 image in this case.

WISE3 emission is shown in the top right panel of Fig.\,\ref{cirrus} over the optical VST image. A quick glance at both Planck857 and WISE3 emission shows there is good correspondence between them, as expected. However, this is not the case for the optical$-$infrared images. Despite its higher resolution, the WISE3 image only matches some areas from the optical image, showing no significant emission nearby CIG\,96.

Following the same procedure as \cite{duc15}, we aimed to trace the Galactic cirrus in the neighbourhood of CIG\,96 with our VST and the WISE3 images. After a careful revision of the reduction and calibration of the VST data, this image shows background emission from a SB level of 28.5\,mag\,arcsec$^{-2}$ and fainter, that partially coincides with the IR emission from the WISE3 image (see Fig.\,\ref{cirrus}, top right panel). We cannot rule out an instrumental origin for some parts of the background emission of the field, yet the partial match between the VST and WISE3 images suggests that most of these structures are not undesired products of a deficient flat-field correction or scattered light but are actual Galactic cirrus, as expected in most latitudes from SB levels of 25\,magarcsec$^{-2}$ or higher \citep{guhathakurta89, cortese10, duc15}. Assuming that the emission belongs to the Galactic cirrus, reaching the very faint SB limit of 28.5\,mag\,arcsec$^{-2}$ or fainter is crucial to assert their detection, implying that shallower images might be missing them. However, we cannot rule out an instrumental origin for the emission with the current data, and further optical and IR images of the same field with the same or fainter SB levels are desirable to confirm the nature of the background structures.

In summary, Planck857 and WISE3 images show no signs of large, diffuse and faint structures over CIG\,96 that might be interfering with our optical detection limits of the galaxy and its structures. However, we confirm a partial correspondence between the WISE3 image and the diffuse optical background emission in the surrounding field. The external (cirrus) or instrumental (flat-field correction) nature of the unmatched structures might be confirmed with further optical images of at least the same SB limit as our VST image. After a careful revision of the VST data processing, we must consider these background structures as a limiting factor to our images, setting the SB limit to 28.4\,mag\,arcsec$^{-2}$.


\section{Discussion of the optical and HI faint structures in the outskirts of CIG\,96}
\label{sec:discussion}

In this section we discuss the implications of the above results with respect to the close environment of CIG\,96 as well as the origins of the different HI and optical features detected.

\subsection{The environment of CIG\,96}
\label{sec:environment}



\begin{figure}
\centering
\includegraphics[width=\hsize]{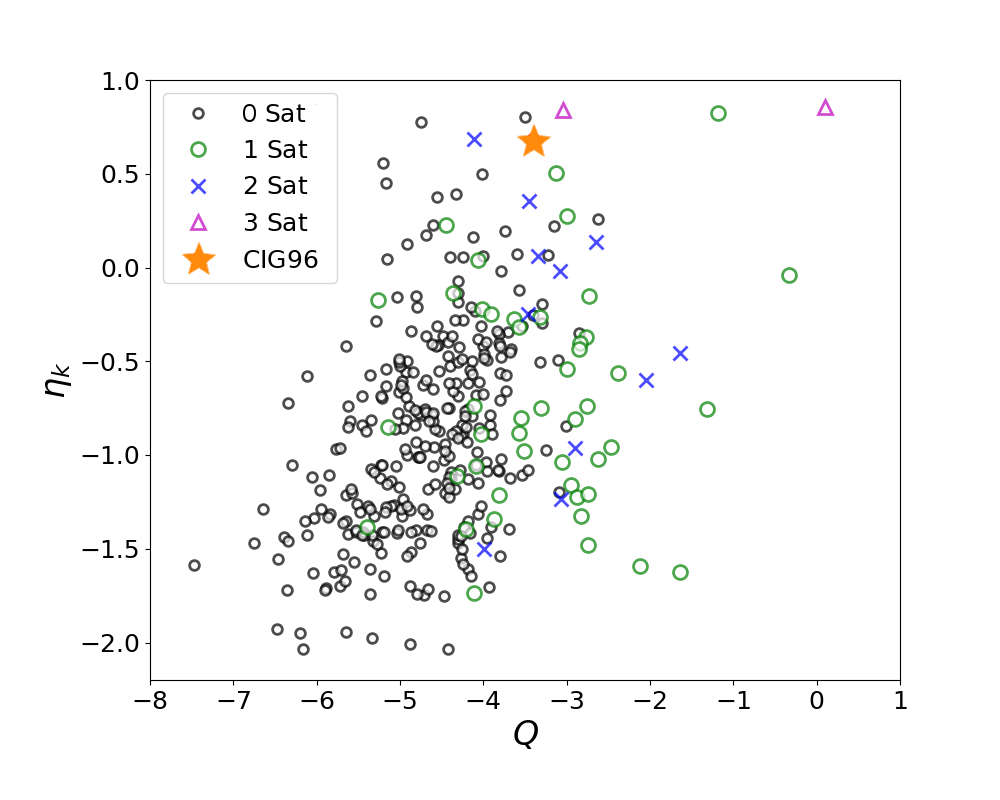}
   \caption{Representation in the isolation parameters of the subsample generated with spectroscopic data of CIG galaxies as calculated by \cite{argudo14} (q.v. Fig. 8 of that work). Lower values of local number density $\eta_k$ and tidal force estimation $Q$ represent higher isolation. The position of CIG\,96 is highlighted.}
      \label{isol_argudo}
\end{figure}



\begingroup
\renewcommand{\arraystretch}{1.3}
\begin{table*}
\caption{Companions of CIG\,96 within 1\,Mpc.\ \tablefootmark{a}}
\label{tablecomp}      
\centering          
\begin{tabular}{c c c c c}     
\hline\hline     
\small{Name} & \small{Apparent B magnitude} & \small{Diameter (major axis)} & \small{Projected distance}          & \small{Central velocity} \\ 
             & \small{(mag)}                & \small{(arcmin / kpc)}        &  \small{(arcmin / kpc / direction)} & \small{(km s$^{-1}$, LSRK)}  \\ 
\hline
NGC\,864\,COM01 	& 16.38 & 0.6 / 3.3  & 15.24\tablefootmark{b} / \ 90\tablefootmark{b} \ / \ \ E  & 1578\tablefootmark{b} \\
HIPASS\,J0217+06	& 16.50  & 1.6 / 9.4  & 40.03 / 235 / NE & 1549 \\
UGC\,01677 		 	& 18.00 & 0.9 / 5.3  & 76.70 / 450 / NW & 1575 \\
UGC\,01670 			& 14.80 & 2.2 / 13.0 & 83.73 / 495 / NW & 1593 \\
UGC\,01803  		& 14.70 & 2.8 / 16.5 & 89.26 / 527 / NE & 1615 \\
\hline
\end{tabular}
\tablefoot{\\
\tablefoottext{a}{Data from the NASA/IPAC Extragalactic Database (NED) unless stated otherwise.} \\
\tablefoottext{b}{This work.} \\
}
\end{table*}
\endgroup
%

The two parameters that quantify the isolation degree of a galaxy from CIG are the local number density, $\eta_{k}$, that accounts for the number of neighbour galaxies within a radius of 0.5 or 1 Mpc (depending on the criteria selected), and the tidal force estimation, $Q$, that quantifies how much the host is affected by its neighbourhood.
According to the NASA/IPAC Extragalactic Database (NED), CIG\,96 has five identified galaxies within a radius of 1\,Mpc: NGC\,864\,COM01 (the aforementioned companion), HIPASS\,J0217+06, UGC\,01677, UGC\,01670 and UGC\,01803. The properties of these neighbours are summarised in Table\,\ref{tablecomp}.

When taking the closest companion, NGC\,864\,COM01, the isolation time\footnote{As defined by \cite{verdes05} and in agreement with the isolation criteria, the isolation time estimates the minimum time a galaxy has been free of interactions with major companions. It is defined as follows: let $D$ be the diameter of CIG\,96; the time that a hypothetical companion of $d=4 \times D$ size needs to cover a distance of 20$\times d$ by travelling at a typical "field" velocity of 150 km\,s$^{-1}$ is 11.5 Gyr, that is, almost the age of the universe, implying no recent interaction. The closest companion NGC\,864\,COM01, travelling at a velocity of 33 km\,s$^{-1}$ with respect to CIG\,96, yields a minimum isolation time of 2.7\,Gyr while the other galaxies have never interacted with CIG\,96.} of CIG\,96 results in $\sim$2.7\,Gyr. The rest of the targets, located farther away, yield isolation times of $\sim$11 Gyr or higher than the age of the Universe, that is, they could not have interacted with CIG\,96 in the past.

We used the methods from two previous works by \cite{verley07b} (hereafter V07) and \cite{argudo13} (hereafter A13) to calculate the isolation parameters of CIG\,96 (see Sect.\,\ref{sec:intro}).
V07 consider the first $k-$th similar sized neighbours within a 500\,kpc radius. In the case of CIG\,96 only the closest two neighbours, i.e. NGC\,864\,COM01 and HIPASS\,J0217+06 are selected and the isolation parameters are $\eta_{k}^{V07} = 0.877$ and $Q_{0.5}^{V07}\,=\,-1.981$.
A13 extend the radius up to 1\,Mpc, that is, they consider the five neighbours mentioned above, and calculate the isolation parameters with photometric and spectroscopic data. The $Q$ parameter of CIG\,96 cannot be calculated with the photometric data because they contain no neighbour that violates the necessary criteria to measure the $Q$ between them and the host. However, the spectroscopic data allow to obtain the isolation parameters for CIG\,96 being $\eta_{k}^{A13}\,=\,0.68$ and $Q^{A13} =\,-\,3.41$. This method is also used by \cite{argudo14}, where they focus on identifying the satellites around host galaxies from CIG, distinguish them from the Large Scale Structure (LSS) and quantify their effect over the host galaxy.

According to the spectroscopic data of A13, the position of CIG\,96 is highlighted in Fig.\,\ref{isol_argudo} where lower values of $\eta_k$ and $Q$ represent higher isolation. 
The remaining galaxies shown for comparison correspond to the subsample of galaxies used in \cite{argudo14} with 0, 1, 2 or 3 dynamically linked satellites (physically bound neighbours, q.v. Fig.\,5 in that work). The five closest companions of CIG\,96 within 1\,Mpc are responsible for raising the $\eta_k$ parameter, whereas the $Q$ value is similar to galaxies with one or two satellites. However, CIG\,96 has no satellite around it. This apparent contradiction reinforces a very relevant point: the isolation parameters are meant for statistical interpretations rather than to understand the environment of a particular galaxy. 
The physical interpretation of these parameters is that the populated environment of CIG\,96 does not contain massive galaxies and, among all the neighbours, the closest companion included, none are affecting the evolution of CIG\,96, as supported by all our observations.
Despite the depth reached with our HI cube and optical images, we detect no signs of gaseous or stellar features between CIG\,96 and its gas$-$rich companion (see Sect.\,\ref{sec:cigcomp} and \ref{sec:velasym}) or any other more distant galaxies. The lack of detections supports the hypothesis of a long-lived isolation of CIG\,96 where its asymmetrical HI distribution (see Fig.\,\ref{mom0}) may come from internal processes rather than an external agent with the exception of an absorbed small companion. HI asymmetries caused by major merger interactions are detectable within 0.4$-$0.7\,Gyr \citep{holwerda11c}, that is, approximately within one or two full rotations of CIG\,96 ($T_{rot}^{CIG\,96}\simeq$ 0.4\,Gyr). However, the high isolation of the galaxy rules out encounters with major companions for at least the last $\sim$2.7\,Gyr.

The Galactic cirrus is observed from a SB level of $\mu_{r\ SDSS} =$\,28.5\,mag\,arcsec$^{-2}$ and fainter, affecting a remarkable area of the field around CIG\,96. Infrared images may play a relevant role for the cirrus identification and future removal, therefore allowing for lower limits in SB to be reached. In this case, the two Planck857 and WISE3 images discussed do not manage to fully trace these structures at the sensitivity and spatial resolution required to remove them from the optical images. Since cirrus cannot be avoided by introducing any changes in the observational and data treatment strategy, they set an observational limit to our optical images, hampering the detection of fainter structures in polluted areas.

After reviewing the DECaLS DR5, a survey of similar characteristics with a SB limit of $\sim$28\,mag\,arcsec$^{-2}$ ($\sim$0.7\,mag\,arcsec$^{-2}$ brighter than our image from VST), we did not spot clear signs of the structures we detect in our VST image. While we cannot fully rule out an instrumental origin, most of the faint structures surrounding CIG\,96 only show up from $\sim$28.5\,mag\,arcsec$^{-2}$ and fainter, suggesting that very faint SB levels are necessary to detect them, possibly being missed by slightly shallower images. We do confirm though the detection of the small stellar feature indicated in the left and central panels of Fig.\,\ref{grpseudocuts} (cyan arrow), a candidate ultra-diffuse galaxy (UDG).

\subsection{Origins of the features of CIG\,96}
\label{sec:feats96}

\textit{HI features}.

For decades it has been known that the HI in spiral galaxies frequently shows asymmetries and lopsidedness \citep{baldwin80, richter94}. However, the origin of such features in the isolated galaxies from CIG is unknown. The asymmetries of CIG\,96 represent an excellent study case and have motivated all the observations and discussion of this work.

As described in Sect.\,\ref{sec:chmap} and \ref{sec:HImoments} and shown in Figs. \ref{figvelchannelwz1}, \ref{mom0} and \ref{pvcuts}, we observe two external HI clouds, the NW and SE HI features, which do not seem to belong to the main HI body of CIG\,96 and a region with a remarkable receding velocity in the outermost SW region.
The NW HI feature is a clumpy, almost continuous cloud spread throughout $\sim$70 km\,s$^{-1}$ and connected to the galaxy approximately along the direction of its minor axis. This connection may be traced down to channels 24 and 25 (1560 and 1570 km\,s$^{-1}$, respectively), where the HI cloud and the galaxy join showing no perturbation in the velocity field.
The SE HI feature consists of a few gaseous clouds located in the same spatial location and spread along $\sim$40 km\,s$^{-1}$. However, the lack of any noticeable perturbation over the outermost HI of CIG\,96 suggest this feature has had little (if any) effect on the eastern side of the galaxy. It is also important to note that neither of these two features show any optical counterpart down to our detection limits.

The most distant HI region located at about $\sim$7$\arcmin$ ($\sim$41 kpc) from the centre along the direction of the major axis (indicated with a cyan arrow in Fig.\,\ref{pvcuts}) shows a receding velocity that is $\sim$30 $-$ 40 km\,s$^{-1}$ above the expected velocity ($\sim$1430 km\,s$^{-1}$) of this region. \cite{espada05} proposed this region as a possible kinematically detached clump from the galaxy. The zeroth moment of our HI cube (Fig.\,\ref{mom0}) and the P/V cuts over the major axis show such a kinematical detachment as well as small and scattered unresolved HI clouds. However, they do not provide evidence of a physical detachment in the SW region (indicated with the cyan arrow in Fig.\,\ref{pvcuts}, top panel), which is larger than any of the small HI clouds nearby. The inspection of the first-moment map suggests the farthest-south region might be warped but this would imply an external agent like a minor merger or the accretion of gas \citep{bournaud05, jog09} that we do not find at our column density limit of $N_{HI}$\,=\,8.9\,$\times$\,10$^{18}$\,cm$^{-2}$. For this reason, we cannot rule out the cold gas accretion below the already low column density reached as a candidate to explain such features.

The P/V profile along the minor axis of CIG\,96 (Fig.\,\ref{pvcuts}) shows emission in a wide gradient of velocities, going up to $+$85 km\,s$^{-1}$ in the receding side and down to $-$145 km\,s$^{-1}$ in the approaching side with respect to its central velocity.
Beam effects may contribute to such dispersion \citep{bosma78} by introducing part of the surrounding emission at different velocities. However, with the current resolution of 28\arcsec$\times$28\arcsec, this effect can only explain dispersions up to approximately $\pm$60 $-$ 70 km\,s$^{-1}$. Counter-rotating gas clouds may also contribute to the high velocity dispersion but we do not detect any signatures of such features in the major axis, where they would likely be visible. An outflow or infall of extraplanar gas of different speeds may also explain the wide range of velocities but we do not detect any signs of either of these in any channel of the map.

\textit{Accretion of cold HI clouds}.

High- and intermediate-velocity HI clouds around the Milky Way may reach masses of 10$^{7} M_{\odot}$. However, the clouds of the Local Group have smaller typical total masses of $M_{HI}^{cloud}\simeq $10$^{5}-$10$^{6} M_{\odot}$ \citep{wakker99} and they are expected to be several orders of magnitude below the total mass of their host galaxies. 

With respect to CIG\,96, from our HI cube we compute a total HI mass of M$_{CIG\ 96}\eqsim$9.77$\times$10$^{9} M_{\odot}$. The two NW and SE HI features have total masses of $\sim3.1$ and 1.6$\times$10$^{6} M_{\odot}$, respectively, close to the detection limit achieved with our HI cube ($0.7 \times 10^{6} M_{\odot}$). The NW HI feature meets the HI mass, spatial distribution, and velocity range criteria to consider it a possible infalling cloud that overlaps with CIG\,96 in channels 23 $-$ 25 (velocities 1550 $-$ 1570 km\,s$^{-1}$).

As discussed by \cite{scott14}, these clouds are not expected to fall on an extension of the rotation curve. This reinforces the idea that the previously discussed SW side of the galaxy is likely to be the warped edge of the HI disc instead of an accreted HI cloud. 

\textit{Pseudo-ring colour, column densities and minor mergers}.

HI is disrupted more easily than the faint optical halo substructures, which may live longer \citep{penarrubia05} than the $\sim$1.5$-$3 Gyr established by the quantified isolation criteria (see Sect.\,\ref{sec:environment}). The external pseudo-ring of CIG\,96 is HI rich except for its southern and NE sides where the gas is scarce. The pseudo-ring colours in the southern side are clearly redder than in the rest of the ring, in particular in the bluer northern and NW sides where SF is taking place in a number of scattered regions according to their blue colours. This colour difference agrees with \textit{GALEX} NUV and FUV results discussed by \cite{espada11a}. 

The star-forming regions with high UV emission match the bluer regions of the N-NW side of the pseudo-ring, and consistently show column densities above 8.5$\times$10$^{20}$ cm$^{-2}$. These are higher than the redder regions which coincide with areas where the HI seems disturbed and show much lower HI column densities, as shown in the bottom panel of Fig.\,\ref{grHI}. 
The $N_{HI}$ is measured with a beam larger than the size of the optical regions. Once resolved, we would expect to achieve an even higher $N_{HI}$ level in the bluest regions, reaching the nominal SF value of $10^{21}$\,cm$^{-2}$.
Also, the anticorrelation observed between the scaled $g-r$ and $N_{HI}$ seems to break in the 70\degree$-$180\degree arc. This is the eastern side of the pseudo-ring where the emission is very diffuse so any correlation or anticorrelation between $g-r$ and $N_{HI}$ in this area is uncertain.

Among the mechanisms that involve the SF enhancement or quenching as well as HI asymmetries are the minor mergers with low-mass dwarf galaxies. 
On the one hand, one or a few recent minor mergers with HI-rich small galaxies (wet mergers) are expected to leave clear HI footprints tracing such events, let alone orbiting stellar structures \citep{martinezdelgado09} or SF enhancement in the areas where the merging occurred: i.e., bluer colours would be expected in the southern and NE areas instead of the redder colours we find. Also, these mergers could have occurred in the more diffuse and incomplete regions of the pseudo-ring but, again, footprints of such events are missing. On the other hand, any minor mergers with one or a few HI-poor companions (dry mergers) might explain the stripped gas as well as the consequent SF quenching (even extinction) of the NE and southern regions. However, we find no optical or gaseous signature in any of these sections and the zeroth and first moments indicate no link between the possible warp in the southern arc of the pseudo-ring. One or various possible encounters with ultra$-$diffuse galaxies (UDG) of large mass$-$to$-$light ratios \citep{vandokkum16} might be responsible for the SF in the pseudo-ring.
The only candidate to UDG lies to the south of the galaxy (see Fig.\,\ref{grpseudocuts} and \ref{radprof}) but we detect no signs of interaction between this feature and the galaxy at the SB level reached with our observations.

The tidal footprints from older minor mergers ($t$ > 0.7 Gyr, \citealt{holwerda11c}) might have disappeared within one or two galaxy revolutions. Should they have left any optical counterparts that might still be visible, they would be expected to be weak and diffuse, such as those detected in the outskirts of other nearby galaxies (with no isolation classification) by \cite{morales18} at similar SB. Except for possible UDG mentioned before, we do not detect any other candidates or tidal streams brighter than our SB limit. However, the limiting SB level of $\mu_{r\ SDSS} =$ 28.5\,mag\,arcsec$^{-2}$ prevents the detection of further potential candidates fainter than this limit, that is, based solely on these optical observations, we may not discard the hypothesis that one or more minor mergers may be responsible for the asymmetries of the galaxy and the colour index variation in the pseudo-ring, nor the possibility that faint optical counterparts to the HI features might exist. Consequently, old minor mergers remain as possible candidates to explain the stellar and gaseous features of CIG\,96.



\textit{Age of the pseudo-ring of CIG\,96}

The blue colour of $g-r$ representations is a good tracer of the age of the stellar population in the optical regime. \cite{schawinski09} propose three different models to discuss age according to the $g-r$ colour observed: model 1 assumes an instantaneous burst of SF with an exponential decay of 100 Myr; model 2 considers an instantaneous burst of SF with instantaneous decay of 10 Myr; lastly, model 3 assumes a constant SFR.

The colour nature of the pseudo-ring of CIG\,96 does not fit with the age estimations of models 1 and 2 because so recent SF would either require nearby companions powerful enough to trigger it, or a transfer of gas from the inner parts towards the pseudo-ring in a lower period of time than the dynamical timescale of the galaxy.
We do not detect signs of any of these requirements, leading us to consider model 3 as a more likely scenario. According to the latter, the continuous and slow ingestion of gas from the central parts of the galaxy into the pseudo-ring might explain a constant SF during at least 1 Gyr.

Additionally, \cite{espada11a} found that the outer parts of CIG\,96 have ultraviolet colours of FUV$-$NUV = 0.1$-$0.2 mag$_{AB}$\,arcsec$^{-2}$. According to this interval FUV$-$NUV, the low-metallicity model proposed by \cite{smith10} fits with the expected lower metallicity in the outskirts of a galaxy. However, it yields ages lower than 100 Myr for the pseudo-ring. We discard the option of a companion since we would expect to detect it further than just its influence on the outskirts.

\textit{Origin of the pseudo-ring of CIG\,96}

The rings located 2$-$2.5 times the radius of the bar of the galaxy are labelled as outer rings (ORs) \citep{buta17}. Outer rings are typical in barred galaxies and their origin is gas accumulation in the outer Lindblad resonance (OLR) \citep{schommer76}.

On the one hand, we find no signatures in either the optical or HI observations that suggest a collisional origin, in agreement with its high isolation level. On the other hand, the OLR is located at a distance slightly beyond twice the bar radius \citep{ath82}. In the case of CIG\,96, the bar and pseudo-ring optical radius are $\sim$2 kpc ($\sim$22$\arcsec$) and $\sim$14\,kpc ($\sim$145$\arcsec$), respectively, that is, the pseudo-ring is located over four times farther than the bar, making such a distance too large to be considered an OLR of the bar, as discussed by \cite{espada05}.

An oval shape of the bright inner disc of the spiral galaxy might be a more reliable source to explain the origin of the pseudo-ring, as discussed briefly by \cite{espada11a}. These non-axisymmetric kinematical features can produce disturbances on the motions of gas clouds located in the outer HI disc, resulting in complete or partial resonance rings \citep{schwarz81, verdes95, buta96}. In agreement with what is expected to be found in these resonance rings, the pseudo-ring of CIG\,96 has a symmetric shape and it is partially defined, showing diffuse optical emission and low HI column densities in regions to the NE and SW.

An apparent misalignment between the galaxy centre and the pseudo-ring might be suggested by external isophotes of CIG\,96 (q.v. Figs. \ref{optvst} and \ref{cig96companion}). This is confirmed by the different elliptical fittings of the pseudo-ring that consistently show its centre lies $\sim$1.2\,kpc ($\sim$12$\arcsec$) to the south of the galaxy centre. However, the diffuse optical eastern side of the pseudo-ring and the high contamination of bright stars around the galaxy prevent a reliable global isophotal fitting analysis and further discussion on this topic.

The nearby stars also prevent us from clearly resolving the two elongated and faint arms or extensions that depart from the north and south of the disc. However, both the UV data analysed by \cite{espada11a} and our CAHA optical images (q.v., Figs. \ref{optvst} and \ref{grpseudocuts}, central panel, yellow crosses) show that the northern and southern extensions connect the disc and the outermost regions of the galaxy where, respectively, they join the western and eastern sides of the pseudo-ring.

\textit{Detection of fainter signatures of interactions}

Up to date, the studies performing deep optical observations of nearby galaxies mostly detect the brightest stellar features located in their outskirts. Standard $\Lambda$-CDM cosmological simulations of galactic halos \citep[e.g.][]{bullock05, johnston08, cooper10} show that a large portion of the debris from old and minor mergers may be fainter than $\sim$30\,mag\,arcsec$^{-2}$. For this reason, we can confirm that we do not detect signatures of these minor interactions in the outskirts of CIG\,96 down to the SB and $N_{HI}$ limits reached with our optical and HI observations, respectively, yet there may be unveiled features lying at fainter SB levels. Currently, for galaxies located close to the Milky Way, the only approach for detecting these very faint remnants of interactions consists of performing stellar density maps of evolved stellar components (e.g. RGB stars) in the halos of these galaxies (e.g. the PISCeS survey and study of Cen A and NGC\,253 \citealt{crnojevic16,crnojevic18}). These studies reach SB limits of 32$-$34\,mag\,arcsec$^{-2}$, yet this technique is not feasible for more distant galaxies like CIG\,96 and farther with the current ground-based telescopes.


\section{Conclusions}
\label{sec:conclusions}

The AMIGA project uses a sample that shows the most symmetric HI integrated profiles when compared to any other sample, even field galaxies. However, some of its members present very asymmetric profiles as well as other features whose origins remain unknown. If large asymmetries are mostly generated by interactions, the lopsidedness of an isolated galaxy such as CIG\,96 should not be observed. 
CIG\,96 is an isolated galaxy of the AMIGA sample that shows a 16\% asymmetry in its HI profile as well as an actively star forming external pseudo-ring detected in the optical, UV and HI regimes.
Our deep optical and HI observations have yielded unprecedented detail of the stellar and gas components of the galaxy and its outskirts. The wide field of view of 1\degree$\times$1\degree of VST telescope and the wavelet-filtered 21-cm data from VLA/EVLA telescope allowed us to reach a maximum SB and column density level of $\mu_{r\ SDSS}$ = 28.7\,mag\,arcsec$^{-2}$ and $N_{HI}$ = 8.9 $\times$ 10$^{18}$ cm$^{-2}$ (5$\sigma$, beam size of 28\arcsec$\times$28\arcsec), respectively. The optical data reveal the detailed structure of the pseudo-ring as well as a gradient in its colour index. Moreover, the HI data show previously undetected features very close to the galaxy. Next, we present the main conclusions of this work:

Down to these limits, we do not find any gaseous or stellar connection between CIG\,96 and any galaxy in its close environment, including its closest, largest, and HI-rich companion NGC\,864\,COM01, located 15.2$\arcmin$ ($\sim$90 kpc) to the east (projected distance) that may be a close to edge-on Sa or Sb galaxy, as the optical and HI properties of the system suggest. Scattered Galactic cirrus shows up from 28.5\,mag\,arcsec$^{-2}$ (SDSS\,$r$ band) in the surroundings of the galaxy and prevents any positive detection of further faint optical features beyond this depth.

We find two low-mass HI features ($\sim$10$^{6} M_{\odot}$) located to the NW and SE edges of the galaxy (the NW and SE HI features). The NW HI feature is visible along a number of immediate channels of the HI cube and depicts a low column density cloud ($N_{HI}^{NW} \simeq$ 6.5$\times$10$^{19}$ cm$^{-2}$) connected with CIG\,96 slightly to the N of its minor axis. We think that the SE HI feature, however, is a series of thin, small and spatially aligned clouds ($N_{HI}^{SE} \simeq$ 4.9$\times$10$^{19}$ cm$^{-2}$) that stand out in the zeroth moment. The individual channels of the HI cube show that the different clouds that compose this feature are not connected to one another and show no direct effect on the immediate gas of the disc edge despite its close proximity.

We find a colour index difference of $g-r \simeq$ 0.4 mag between two sides of the partially complete pseudo-ring (PA$_{redder}$ = 70$\degree- $258$\degree$ and PA$_{bluer}$ = 258$\degree -$38$\degree$) that cannot be assigned to any instrumental effect. 
No environmental cause (external gas accretion or minor merger) has been identified in our data as to explain such a change in the colour index. The outermost star-forming regions detected with NUV and FUV images from GALEX coherently match the blue regions of the pseudo-ring, which also show $N_{HI}$ values close to 10$^{21}$ cm$^{-2}$, the nominal SF value. The cause for the higher concentration of $N_{HI}$ in certain areas of the pseudo-ring is still to be found.

We have reviewed different SF models based on the FUV$-$NUV and $g-r$ colours to determine the approximate age of the pseudo-ring of CIG\,96. We may discard a short lived pseudo-ring ($\sim$100 Myr or younger) caused by a very recent encounter with either a similar-sized galaxy (the isolation criteria discard them) or a small galaxy (we would expect to see the merger remains); instead, they favour an older pseudo-ring ($\sim$1 Gyr).

Despite the fact that bars are usually relevant candidates to play a critical role in the secular evolution of the outskirts of a galaxy by leading the matter to concentrate in the OLR, such is not the case for CIG\,96. First, we do not find any significant matter concentration in the OLR of CIG\,96 and, second, the pseudo-ring is located at almost double the radius of the expected location of the OLR based on the bar size. For these reasons, we cannot consider a bar-driven accumulation of matter in the OLR as the pseudo-ring origin. Either an oval distortion or old, elongated arms expelled from the inner parts of the disc are more fitting explanations of the origin of the faint, distant (from the galaxy centre) and circular pseudo-ring of CIG\,96. 
Star-forming regions are expected in secular evolution and they may be triggered by external factors such as encounters with smaller and fainter infalling HI clouds. However, we do not find evidence of any external event that may explain the star-forming regions of the pseudo-ring of CIG\,96. Their origin remains unknown.

The lack of any remarkable tidal features or other stellar components leads us to consider that, on the one hand, any major encounter with similar sized galaxies must have never occurred, as guaranteed by the isolation times and criteria (see Sect.\,\ref{sec:environment}); on the other hand, any possible minor merger must have taken place before the last two revolutions of CIG\,96 ($t$\,>\,0.8\,Gyr, approximately), allowing the footprints of such encounters to disappear within such a time.

\begin{acknowledgements} 

P.R.M. and all the coauthors thank the referee for the careful reading and valuable suggestions provided in the report, which have helped to improve this paper significantly. P.R.M. is funded by the project AMIGA4GAS project and the FPI Grant AYA2011-30491-C02-01 by the Ministry of Economy and Competitiveness of Spain. P.R.M., L.V.M., J.B.H., M.J., M.F.L., S.S.E., J.G.S. acknowledge support from the grant AYA2015-65973-C3-1-R (MINECO/FEDER, UE). P.R.M acknowdledges V\'ictor Terr\'on for his assistance during the calibration of the optical data. P.R.M. and the rest of the coauthors acknowledge Tom Jarrett for providing the WISE image used in this work; I.Trujillo for the discussion with respect to the deep optical observations and techniques; Monika Petr-Gotzens (ESO) for the help, follow$-$up and good advices provided with the VST observations and Enrique P\'erez for his help with the dynamical mass estimations. L.V.M. acknowledges discussions with Alberto Fern\'andez$-$Soto about the CAHA1.23m data. J.I.P. acknowledges financial support from the Spanish Ministerio de Econom\'{\i}a y Competitividad under grant AYA2013-47742-C4-1-P, and from Junta de Andaluc\'{\i}a  Excellence Project PEX2011-FQM-7058. This publication makes use of data products from the Wide-field Infrared Survey Explorer, which is a joint project of the University of California, Los Angeles, and the Jet Propulsion Laboratory/California Institute of Technology, funded by the National Aeronautics and Space Administration. The National Radio Astronomy Observatory (NRAO) Karl G. Jansky Very Large Array (VLA) is a facility of the National Science Foundation operated under cooperative agreement by Associated Universities, Inc.. This research made use of Astropy, a community-developed core Python package for Astronomy \citep[][]{astropy13, astropy18} and of APLpy, an open-source plotting package for Python \citep{robi12}. We used the NASA/IPAC Extragalactic Database (NED), operated by the Jet Propulsion Laboratory, California Institute of Technology, under contract with the National Aeronautics and Space Administration. IRAF is distributed by the National Optical Astronomy Observatories, which are operated by the Association of Universities for Research in Astronomy, Inc., under cooperative agreement with the National Science Foundation. 

\end{acknowledgements}


\bibliography{refs.bib}

\end{document}